\documentclass[apj]{emulateapj}
\pdfoutput=1
\usepackage{amsmath}
\usepackage{ulem}
\usepackage{CJK}
\usepackage{graphicx,subfigure}
\usepackage{amsmath}
\usepackage{rotating}
\usepackage{epsfig}
\usepackage{epstopdf}
\usepackage{booktabs}

\newcommand\bld[1]{\mbox{\boldmath $#1$}}

\newcommand{\del}{\partial}

\newcommand{\be}{\begin{equation}}
\newcommand{\ee}{\end{equation}}
\newcommand{\bea}{\begin{eqnarray}}
\newcommand{\eea}{\end{eqnarray}}

\shorttitle{Collision-induced magnetic reconnections \& energy dissipations}
\shortauthors{Deng, Li, Zhang \& Li}
\begin{document}
\begin{CJK*}{UTF8}{gbsn}
\title{Relativistic MHD simulations of collision-induced magnetic dissipation in Poynting-flux-dominated jets/outflows}
\author{Wei Deng (邓巍)\altaffilmark{1,}\altaffilmark{2}, Hui Li (李晖)\altaffilmark{2}, Bing Zhang (张冰)\altaffilmark{1}, Shengtai Li (李胜台)\altaffilmark{2}}
\altaffiltext{1}{Department of Physics and Astronomy, University of Nevada Las Vegas, Las Vegas, NV 89154, USA; deng@physics.unlv.edu; zhang@physics.unlv.edu}
\altaffiltext{2}{Los Alamos National Laboratory, Los Alamos, NM 87545, USA; hli@lanl.gov; sli@lanl.gov}
\altaffiltext{}{LA-UR-15-20564}

\begin{abstract}

We perform 3D relativistic ideal MHD simulations to study the collisions between high-$\sigma$ (Poynting-flux-dominated) blobs which contain both poloidal and toroidal magnetic field components. This is meant to mimic the interactions inside a highly variable Poynting-flux-dominated jet. We discover a significant electromagnetic field (EMF) energy dissipation with an Alfv\'enic rate with the efficiency around 35\%.
Detailed analyses show that this dissipation is mostly facilitated by the collision-induced magnetic reconnection. Additional resolution and parameter studies show a robust result that the relative EMF energy dissipation efficiency is nearly independent of the numerical resolution or most physical parameters in the relevant parameter range.
The reconnection outflows in our simulation can potentially form the multi-orientation relativistic mini-jets as needed for several analytical models. We also find a linear relationship between the $\sigma$ values before and after the major EMF energy dissipation process.
Our results give support to the proposed astrophysical models that invoke significant magnetic energy dissipation in Poynting-flux-dominated jets, such as the internal collision-induced magnetic reconnection and turbulence (ICMART) model for GRBs, and reconnection triggered mini-jets model for AGNs.
\end{abstract}

\section{Introduction}\label{sec:intro}

The energy composition in the jet/outflow of astrophysical systems is an important and fundamental question, since it leads to significant differences for the subsequent energy dissipation process, particle acceleration mechanism, radiation spectrum and light curve, polarization behavior, neutrino emission luminosity, and so on. Generally speaking, jets can be separated into two types depending on their energy composition: Poynting-flux-dominated (PFD) ($\sigma>>1$) and matter-flux-dominated (MFD) ($\sigma<<1$), where $\sigma$ is the magnetization parameter defined as the ratio between the electromagnetic field (EMF) energy flux to the plasma matter energy flux. 

Many independent observations from Gamma-Ray Bursts (GRBs), Active Galactic Nuclei (AGNs), micro-quasars, and Crab nebula give strong hints of the PFD outflows at least for some events. Several theoretical models have been proposed within the framework of PFD jets/outflows to interpret the observations.

In the field of GRBs, evidence of PFD jets has been collected independently in several directions. First, a prominent thermal emission component as expected in the fireball-internal-shock model \citep[e.g.][]{meszaros00} has been seen only in a small fraction of GRBs (e.g. GRB 090902B, \citealt{ryde10,zhang11}). The majority of GRBs either show no evidence of a thermal component or a weak, sub-dominant thermal component \citep[e.g.][]{abdo09a,guiriec11,axelsson12}. These GRBs require that the GRB central engine is highly magnetized, and jet is still PFD at the emission site \citep{zhangpeer09,GaoZhang14}. Next, strong linear polarization was discovered during the prompt gamma-ray emission phase for some GRBs \citep{yonetoku11,yonetoku12}, and during the reverse-shock-dominated early optical afterglow emission phase for some others \citep{steele09,mundell13}, which hint at the existence of globally ordered magnetic fields in the jet. Furthermore, strong PeV neutrino emission as predicted by the MFD models has not been observed from GRBs so far \citep{abbasi12}, which is consistent with the expectation of the PFD models \citep{zhangkumar13}. Finally, the MFD internal shock (IS) model for GRBs also suffers some criticisms, such as low energy dissipation efficiency \citep{panaitescu99,kumar99}, electron fast cooling \citep{ghisellini00}, the electron number excess \citep{bykov96,daigne98,shen09}, and inconsistency with some empirical (Amati/Yonetoku) relations \citep{zhangmeszaros02c,liang10}. \cite{ZhangYan11} proposed a novel PFD outflow model named as ``the Internal-Collision-induced MAgnetic Reconnection and Turbulence (ICMART)", which can potentially keep the merits of the IS model but alleviate the criticisms faced by the IS model mentioned above. The main idea of the ICMART model is that the GRB jets are Poynting-flux dominated. The Poynting flux is catastrophically discharged at a relatively large distance (e.g. $10^{15}$ cm) from the central engine through collision-induced magnetic reconnection. The magnetic energy is converted to particle energy and radiation efficiently, leading to a very high radiation efficiency as demanded by the GRB data \citep{panaitescu02,zhang07}. A PFD jet has less leptons than the MFD model so that the electron excess problem is avoided. A large emission radius favors a moderately fast cooling, which can account for the right low-energy spectral index observed in GRBs \citep{uhm14}. It also gives a natural explanation of the seconds-duration of ``slow variability component''observed in GRBs \citep{Gao12}. The rapid ``fast variability component'' can be interpreted within this scenario as mini-jets due to locally Lorentz boosted regions (see also \citealt{lyutikov03,narayan09}\footnote{\cite{lyutikov03} and \cite{narayan09} proposed that GRB variability is a consequence of mini-jets due to relativistic outflow from reconnection or relativitic turbulence. There is no simple explanation to the observed slow variability component in these models. \cite{ZhangYan11} attributed the two variability components (slow and fast) as due to central engine activity and mini-jets, respectively. Monte Carlo simulations by \cite{ZhangZhang14} showed that the ICMART model can indeed reproduce the observed GRB light curves.}). It is speculated that turbulent reconnection in a moderately high-$\sigma$ flow can give rise to relativistic motion of mini-jets within the bulk relativistic motion of the jets.

For AGNs, observations show fast variable TeV flares of two blazars (Mrk 501 and PKS 2155-304) \citep{aharonian07,albert07}. The light-crossing time is even shorter than the event horizon size of the black holes, so that emission must come from a small local region. The derived Lorentz factor in the emission region should be larger than 50 \citep{begelman08,mastichiadis08}. This value is much larger than the observed Lorentz factor of the bulk motion of the global jet, which is generally smaller than 10 \citep{giroletti04,piner04}. To interpret these observations, \cite{Giannios09} proposed a ``jets in a jet" model, which considers that some mini-jets are generated by local reconnection outflows in a global PFD jet. The mini-jets can give extra Lorentz boosting and particle acceleration to generate the observed TeV photons around these local reconnection regions with fast variability. Even though \cite{Giannios09} did not specify the mechanism of magnetic dissipation, observations of AGN jets reveal bright knots that are consistent with internal interactions within the jet. Within the PDF jet scenario, ICMART processes similar to what are envisaged in GRB jets may also play a role.

Another related astrophysical phenomenon is $\gamma$-ray flares observed from the Crab nebula. Monte carlo simulations suggest that the bright $\gamma$-ray flares and fluctuations in longer time scales can be understood within the framework that there are many mini-jets with a wide distribution of size and Lorentz factor within the PFD outflow of the pulsar. The flares correspond to the epochs when some bright mini-jets beaming towards earth \citep{yuan11}. The observations suggest that similar physical processes as those operating in GRBs and AGNs may be playing a role in the Crab nebula.

In another front, recent Partical-In-Cell (PIC) simulations \citep{sironi14,guo14} show that reconnection under high-$\sigma$ condition can efficiently accelerate thermal particles to form a non-thermal power-law population of the particles. This gives a good support to the above PFD models from the particle acceleration point of view.

The models discussed above for different astrophysical systems share some common physical processes, such as efficient magnetic energy dissipation in the PFD outflow/jet, mini-jets generated by the relativistic outflows due to local reconnections, particle acceleration in the reconnection region, and production of the non-thermal emission. Although these models show great potential to interpret the observations and overcome the criticisms in the traditional MFD models, some important ingradients of the models are still of a speculative nature. Detailed numerical simulations are needed to give a solid footing to these models.

From the morphologic point of view, jets/outflows can be categorized into two types: continuous and episodic. Theoretically, episodic jets can be formed either from a highly variable central engine with variable accretion rate; or disruption of a continuous jet by screw or kink instabilities \citep{li00,mizuno09}; or from a MHD erruption process similar to solar coronal mass ejection \citep{yuan09,YuanZhang12}. Observationally, episodic jets or knots in jets have been observed in many X-ray binaries \citep{mirabel94,hjellming95,fender2004} and AGNs \citep{marscher02,chat09,doi2011}. Rapid variabilities observed in GRBs also point towards highly episodic jets \citep{rees94,paczynski94}. As a result, studying interactions or collisions between magnetic blobs or shells is of great interest.

In this paper, we perform detailed numerical simulations on the global properties of collisions between high-$\sigma$ blobs, as envisaged in the ICMART model of GRBs \citep{ZhangYan11}. In Section \S \ref{sec:setup}, we give a brief introduction of our 3D relativistic MHD code and the simulation setup. In Section \S \ref{sec:example}, we present an example simulation case to show the key results, and perform a detailed analysis and resolution study. We then expand our simulations on two-blob collisions in Section \S \ref{sec:para} to a large parameter space and discuss how different parameters affect the simulations results. In Section \S \ref{sec:4blobs}, we show preliminary results for multiple collisions among four high-$\sigma$ blobs. We summarize our results in Section \S \ref{sec:summary} and discuss the implications of our simulation results on some high energy astrophysical systems, such as GRBs and AGNs.

\section{Numerical method and problem setup}\label{sec:setup}

\subsection{Code introduction}

We use a 3D special relativistic MHD (SRMHD) code which solves the conservative form of the ideal MHD equations using higher-order Godunov-type finite-volume methods. This code is a development version of the ``LA-COMPASS" MHD code which was first developed by \cite{LiLi03} at Los Alamos National Laboratory.
The equations solved in the code are:
\begin{align}
&\frac{\partial(\Gamma \rho)}{\partial t}+\nabla \cdot (\Gamma \rho \bld{V}) = 0,\\
&\frac{\partial}{\partial t}(\frac{\Gamma^2 h}{c^2}\bld{V}+\frac{\bld{E}\times \bld{B}}{4 \pi c})+
\nabla \cdot[\frac{\Gamma^2 h}{c^2}\bld{V}\otimes \bld{V}+(p+\frac{B^2+E^2}{8 \pi})\bld{I} \nonumber \\
&-\frac{\bld{E}\otimes \bld{E}+\bld{B}\otimes \bld{B}}{4 \pi}] = 0, \\
&\frac{\partial}{\partial t}(\Gamma^2 h-p-\Gamma \rho c^2 + \frac{B^2+E^2}{8 \pi}) + \nabla \cdot[(\Gamma^2 h-\Gamma \rho c^2)\bld{V} \nonumber \\
&+\frac{c}{4 \pi}\bld{E}\times \bld{B}]=0,\\
&\frac{\partial \bld{B}}{\partial t}+ c \nabla \times \bld{E}=0,\\
&\bld{E}=-\frac{\bld{V}}{c}\times \bld{B},
\end{align}
where $\Gamma$, $\rho$, $h$, $P$ are the Lorentz factor, rest mass density, relativistic enthalpy, and gas pressure, respectively, $\bld{V}$, $\bld{E}$, $\bld{B}$ are the vectors of fluid velocity, electric field, and magnetic field, respectively, and the symbol ``$\otimes$'' denotes tensor product. We also use the ideal gas equation of state: $p=(\hat{\gamma}-1)u$, where $\hat{\gamma}$ and $u$ are the adiabatic index and the internal energy density, respectively. 

We use HLL flux with the piecewise parabolic reconstruction method to solve the Riemann problem \citep{colella84}, and use the constrained transport (CT) method \citep{balsara99,guan14} to ensure $\nabla \cdot \bld{B}=0$. We use the Cartesian coordinates $(x, y, z)$ in our simulations.

\subsection{Problem set up}

We envisage that the central engine of GRBs or AGNs launch a Poynting-flux-dominated jet/outflow. As discussed in Sect. 1,  episodic jets are preferred from observational data. Even if the jet may be overall continuous, it is very likely non-uniform internally and may form many knots in the jet, where a much larger amount of EMF energy ($E_{\rm em}$) is concentrated compared with other sparse regions in the jet. We can simplify the knots of the jet/outflow as many quasi-isolated magnetic blobs with both poloidal and toroidal field components. Due to the intrinsic erratic behavior at the central engine, different magnetic blobs may have different velocities at the emission region, so that multiple collisions are very likely to happen among different blobs. Due to the ultra relativistic motion of the jet, the relative velocities between different blobs can easily become relativistic.

In our simulation domain, we use the model from \cite{Li06} to initialize the magnetic field configuration. The equations are introduced in the cylindrical coordinates $(r, \phi, z)$, and we will transfer them to the Cartesian coordinates $(x, y, z)$ in our simulations. from the center ($r=0$) of each blobs, the field is assumed to be axisymmetric with the poloidal flux function $\Phi$ as
\begin{equation}
\Phi(r,z) = B_{\rm b,0} r^2\exp \left(-\frac{r^2+z^2}{r^2_0}\right),
\label{equ:phi}
\end{equation}
and the relationship between $\Phi(r,z)$ and the $\phi$ component of the vector potential $A_{\phi}$ is 
$\Phi(r,z) = r A_{\phi}$. $B_{\rm b,0}$ and $r_0$ are the normalization factor for the magnetic strength and characteristic radius of the magnetic blob, respectively.
One can then calculate the $r-$ and $z-$ components of the poloidal field
\begin{equation}
B_{r} = -\frac{1}{r}\frac{\del \Phi}{\del z} = 2B_{\rm b,0}\frac{zr}{r^2_{0}} \exp \left(-\frac{r^2+z^2}{r^2_{0}}\right),
\label{equ:Br}
\end{equation}
and
\begin{equation}
B_{z} = \frac{1}{r}\frac{\del \Phi}{\del r} = 2B_{\rm b,0} \left(1-\frac{r^2}{r^2_{0}}\right)\exp\left(-\frac{r^2+z^2}{r^2_{0}}\right).\label{equ:Bz}
\end{equation}

The poloidal field is closed and keeps the net global poloidal flux as zero. The toroidal field configuration is motivated by considering the black hole accretion disk system as a ``dynamo", which shears the poloidal flux to form the toroidal flux from the rotation. The toroidal component of the magnetic field therefore has the form
\begin{equation}
B_{\phi}  = \frac{\alpha\Phi}{r_0 r} = B_{\rm b,0} \alpha \frac{r}{r_0}\exp\left(-\frac{r^2+z^2}{r^2_{0}}\right)~.
\label{equ:B_phi}
\end{equation}
Here the parameter $\alpha$ controls the toroidal-to-poloidal flux ratio. \cite{Li06} showed that when $\alpha \sim 3$, the two flux components are roughly equal with each other. We set $\alpha=3$ for our example simulation, and explore a larger value of $\alpha$ in Section \S\ref{subsec:alpha}. We choose the comoving center-of-mass frame of the blobs as our simulation frame. The direction of velocity is along $Z$-axis with a profile
\begin{equation}
V_z = \left \{
 \begin{array}{ll}
 V_{\rm b,z}, & (r\leq r_0),\\ \\
 V_{\rm b,z} \exp\left(-\left(\frac{r-r_0}{r_0/2}\right)^2\right), & (r>r_0),
 \end{array}
 \right.
\label{equ:vel}
\end{equation}
where $V_{\rm b,z}$ is a constant value which can be either positive or negative corresponding to $+Z$ or $-Z$ direction of the velocity. We also set a uniform gas pressure value ($P$) both inside and outside the blobs. The value of $P$ is much smaller than the initial magnetic energy density of the blobs.

For the density profile, we first define a constant initial value of the blob magnetization parameter around the central region of the blobs:
\begin{equation} 
\sigma_{\rm b,i}= \frac{E_{\rm em}}{\Gamma^2 h},
\end{equation} 
where $h=\rho c^2+\hat{\gamma}P/(\hat{\gamma}-1)$ is the specific enthalpy defined in the fluid's comoving frame, $\rho$ is the rest mass density, $P$ is the gas pressure introduced above, $\hat{\gamma}$ is the adiabatic index, $\Gamma$ is the bulk Lorentz factor calculated by the velocity profile introduced above, and $E_{\rm em}$ is the EMF energy density calculated by $E_{\rm em}=(\bld{B}^2+\bld{E}^2)/8 \pi$ from the magnetic field profile introduced above. The density profile is therefore
\begin{equation}
\rho = \left \{
 \begin{array}{ll}
 \frac{1}{c^2}\left(\frac{E_{\rm em}}{\Gamma^2 \sigma_{\rm b,i}} - \frac{\hat{\gamma}P}{\hat{\gamma}-1}\right), & (r \leq r_0 ~~ {\rm and} ~~ \rho > \rho_{\rm bkg}),\\ \\
 \rho_{\rm bkg}, & (r > r_0 ~~ {\rm or} ~~ \rho \leq \rho_{\rm bkg}),
 \end{array}
 \right.
\label{equ:vel}
\end{equation}
where $\rho_{\rm bkg}$ is a constant parameter to control the uniform background mass density.

We also introduce two position-control parameters $z_d$ and $x_s$. For a collision between two blobs, the center of the two blobs are located at $(x_1,y,z_1)$ and $(x_2,y,z_2)$, so $z_d=|z_1-z_2|$ is the initial distance between the center of the two blobs in $Z$ direction, and $x_s=|x_1-x_2|$ is the initial misalignment between the center of the two blobs in $X$ direction due to the possible misalignment of the blobs. The $Y$ coordinate is the same for both of them.

In Table \ref{tab:t2}, we give the normalization relationship between the code units and the physical units.  There are only three free parameters, $L_0$, $B_0$ and $c$ to control the normalization of the entire system. Defining different physical values of $L_0$ and $B_0$, we can normalize the simulation system to different environments and problems. In Table \ref{tab:t2}, we also list two sets of example typical values to show the way of the normalization.
In the rest of paper, all the parameters are given using code units. We keep $r_0=1.0$ for all the following simulations. In addition, we use $\hat{\gamma}=5/3$ in most of the simulations, since most of the regimes are mildly relativistic. This may not always be true, especially in the regions of reconnection outflows, so in Section \S \ref{subsec:adi} we also try $\hat{\gamma}=4/3$ to test the difference.

\begin{table*}[!htb]
\centering
\caption{The normalization factors between physical units and code units.}
\begin{tabular}{ccccccc}
\toprule
Parameters: & Length & Velocity & Time & Magnetic field & Pressure & Density\\
Code units: & 1 & 1 & 1 & 1 & 1 & 1\\
Normalization factors: & $L_0$ & $c$ & $L_0/c$ & $B_0$ & $B_0^2$ & $B_0^2/c^2$\\
Typical values 1: & $10^{12}$ cm & $3\times 10^{10}$ cm/s & $33$ s & $10^{3}$ G & $10^{6}$ Ba & $1.1\times 10^{-15}$ g/cm$^3$\\
Typical values 2: & $10^{13}$ cm & $3\times 10^{10}$ cm/s & $333$ s & $10$ G & $10^{2}$ Ba & $1.1\times 10^{-19}$ g/cm$^3$\\
\bottomrule
\end{tabular}
\label{tab:t2}
\end{table*}

\section{An example case}\label{sec:example}

In this section, we show a series of detailed analyses based on one example simulation case. We focus on the following aspects: the evolution of magnetic energy to address the efficiency of magnetic energy dissipation, the details of the collision process, the properties of magnetic reconnection and outflows, and the numerical resolution effects. We reveal significant collision-induced reconnection events with a remarkable efficiency around 35\%, which is resolution insensitive. The outflow properties of reconnection events indicate the potential capability to generate super-Alfv\'enic relativistic mini-jets.

\subsection{Initial parameters}

\begin{figure*}[!htb]
\plotone{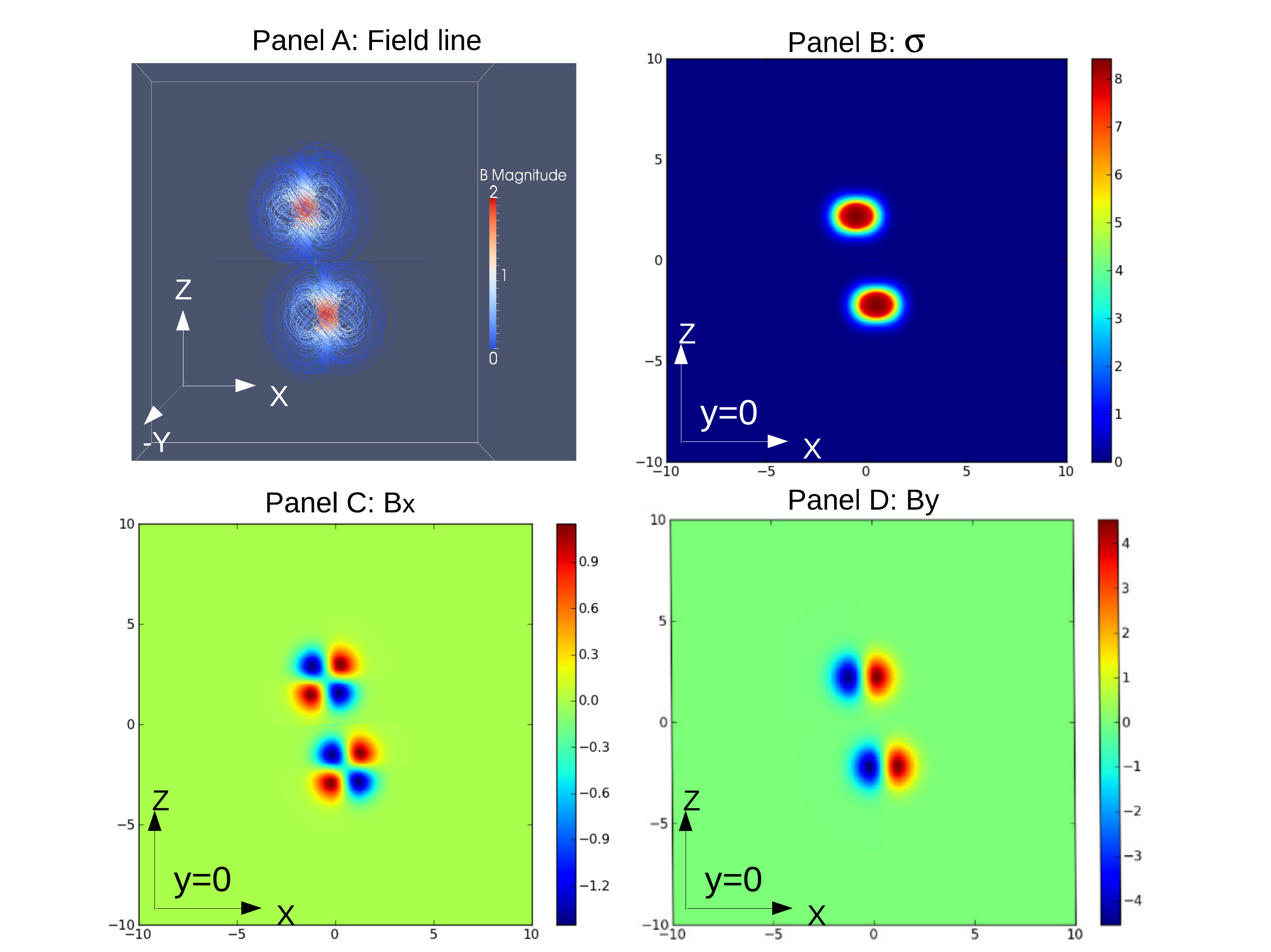}
\caption{Several manifestations of the initial magnetic field configuration cut in the blob-center plane in the example simulation. \textit{Panel A}: The initial 3D field line profile viewed along the $Y$ direction. The color contour denotes the value of $B/\sqrt{4\pi}$; \textit{Panel B}: The 2D contour cut of the initial $\sigma$ profile in the $XZ$ plane ($y=0$);  \textit{Panel C}: The 2D contour cut of the $x$-component of the initial magnetic field strength in the $XZ$ plane ($y=0$); \textit{Panel D}: The 2D contour cut of the $y$-component of the initial magnetic field strength in the $XZ$ plane ($y=0$).}
\label{fig:initial_cut}
\end{figure*}

The initial parameters of the example run are listed in Table \ref{tab:t1}. We consider two identical blobs with initial magnetization parameter $\sigma_{\rm b,i} = 8$ separated by $z_d = 4.4$, with an $X$-direction offset 1.0. The two blobs move in opposite directions in $Z$ direction with an initial center speed $V_{\rm b,z}=0.3$ c. The background pressure and density are $P=10^{-2}$ and $\rho_{\rm bkg}=10^{-1}$, respectively. In order to clearly show the initial magnetic field configuration of the blobs, in Figure \ref{fig:initial_cut} we show a $y=0$ slice (cut through the blob centers) of the profiles of several parameters: projected field line configuration (panel A), $\sigma$ distribution (panel B), $B_x$ (panel C), and $B_y$ (panel D).

For this example run, the 3D box size is chosen as $20^3$ from -10 to +10 in each dimension, which means that the position (x,y,z)=(0,0,0) corresponds to the center of the box. And the resolution is chosen as $1024^3$.

\begin{figure}[!htb]
\plotone{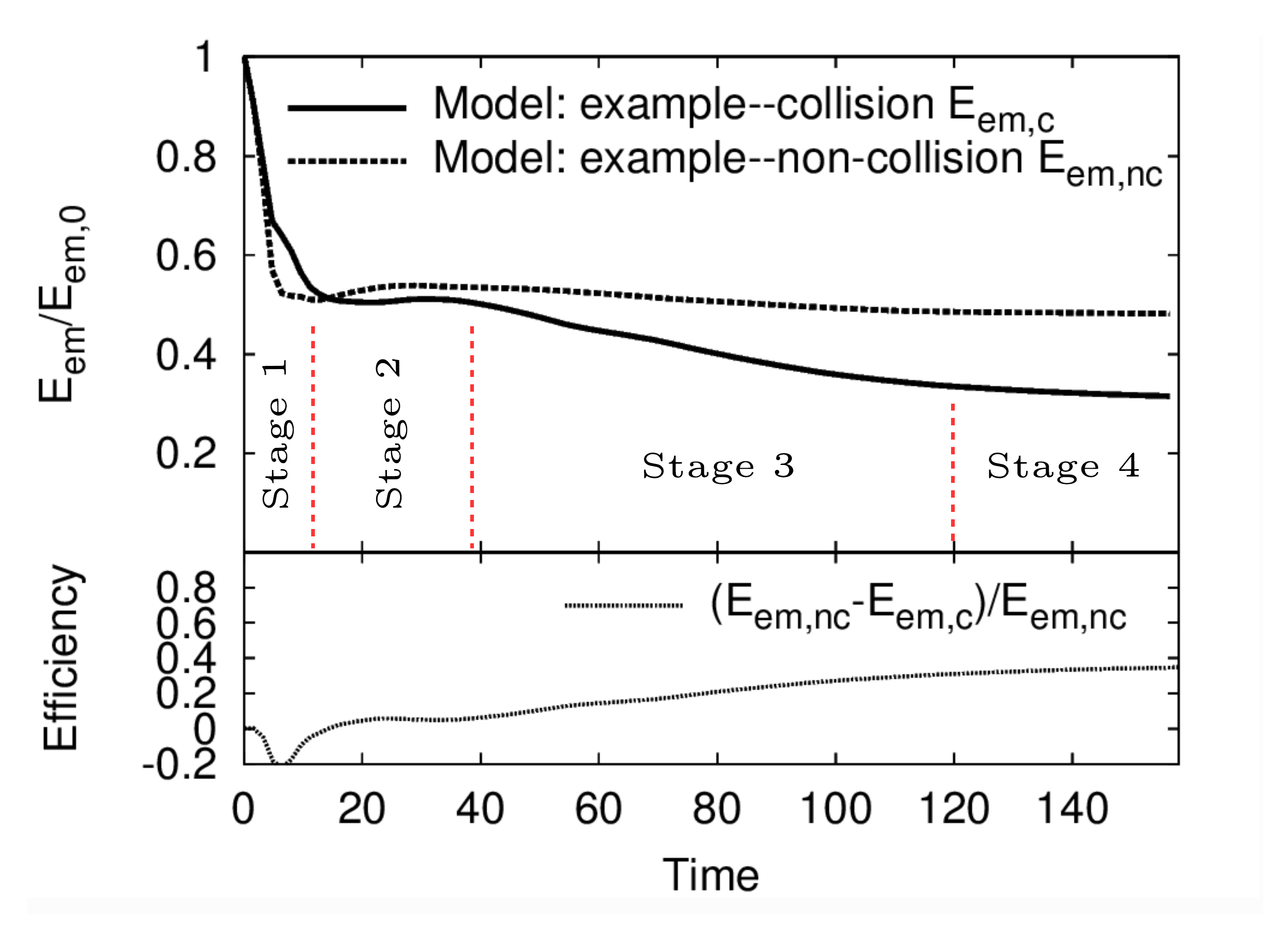}
\caption{\textit{Upper panel}: The Poynting flux energy ($E_{\rm em}$) evolution of the example simulation case. Dashed line denotes the non-collision case, which serves as the reference for additional magnetic dissipation. Solid line denotes the case of collision between two blobs. \textit{lower panel}: Ratio calculated by $(E_{\rm em,nc}-E_{\rm em,c})/E_{\rm em,nc}$ to show the additional $E_{\rm em}$ dissipation efficiency triggered by the collision-induced processes.}
\label{fig:f1}
\end{figure}

\begin{table}[!htb]
\centering
\caption{The initial parameters for the example simulation.}
\begin{tabular}{cccccccc}
\toprule
$\sigma_{\rm b,i}$ & $B_{\rm b,0}$ & $\alpha$ & $\left|V_{\rm b,z}\right|$ & $P$ & $\rho_{\rm bkg}$ & $z_d$ & $x_s$ \\
\midrule
8 & $\sqrt{4\pi}$ & 3 & 0.3c & $10^{-2}$ & $10^{-1}$ & 4.4 & 1.0\\
\bottomrule
\end{tabular}
\label{tab:t1}
\end{table}

\begin{figure*}[!htb]
\plotone{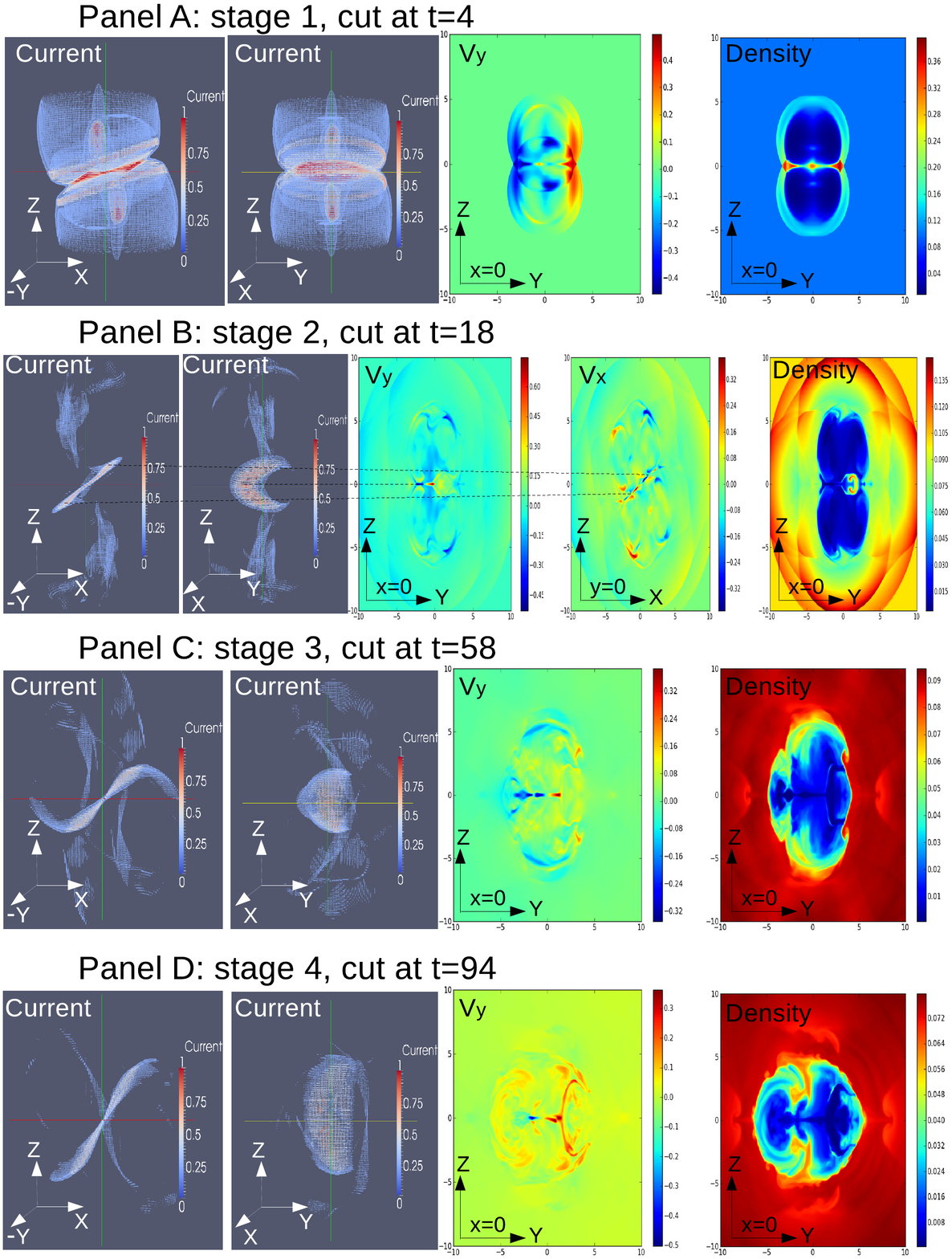}
\caption{The representative cuts of current, velocity and density for the different evolution stages corresponding to Figure \ref{fig:f1}. Panel A corresponds to the initial ``self adjustment" phase; Panel B corresponds to the following ``plateau" phase; Panel C \& D correspond to the ``normal decay" phase. The last quasi-steady phase has no obvious feature, so we do not draw cuts for that stage. For each panel, the cuts from left to right are the 3D current contour plot viewed from $Y$-axis, the 3D current contour plot viewed from $X$-axis, the 2D contour cut of the $y$-component of the outflow velocity ($V_y$) in the $YZ$-plane (x=0) corresponding to the current plot, and the 2D contour cut of the rest mass density in the $YZ$-plane (x=0), respectively. In Panel B, we add an additional 2D contour cut of the $x$-component of outflow velocity ($V_x$) in the $XZ$-plane (y=0) to show the existence of multiple directions of the outflows.}
\label{fig:stages}
\end{figure*}

\subsection{Energy evolution analysis}\label{subsec:energy_evo}

Since the initial magnetic configuration is not in complete force balance (between the internal magnetic pressure and the background gas pressure), the blobs would quickly expand and evolve into a quasi-steady phase, forming a quasi-force balance between the gas pressure and magnetic pressure. During this process, a fraction of EMF energy $E_{\rm em}$ is converted to thermal and kinetic energy due to magnetic field relaxation. So before performing a collision simulation, we first simulate the blob evolution of non-collision case to quantify the EMF energy level in the quasi-steady phase. This would serve as the reference value to be compared with the collision case in which additional EMF energy drop is expected due to additional magnetic dissipation.

The upper panel of Figure \ref{fig:f1} shows the evolution of the blob electromagnetic energy $E_{\rm em}$ as a function of time (normalized to the initial value $E_{\rm em,0}$). The dashed line shows the evolution in the non-collision case. There is a significant drop of $E_{\rm em}$ before $t\sim 6$, which is due to the magnetic field relaxation during the process of establishing a force balance between the outward magnetic pressure force and the inward gas pressure force. After the balance is established, $E_{\rm em}$ is nearly constant and enters a quasi-steady phase, which can be used as the reference energy level without collision.

Next, we simulate the collision case between two high-$\sigma$ blobs. The initial parameters for these two blobs are the same as the non-collision case. The $E_{\rm em}$ evolution of the two blobs with collision is shown as the solid line in the upper panel of Figure \ref{fig:f1}. The efficiency ($\eta$) of $E_{\rm em}$ energy dissipation due to collision-induced process can be calculated by
\begin{equation}
\eta=\frac{E_{\rm em,nc}-E_{\rm em ,c}}{E_{\rm em,nc}},
\label{equ:phi}
\end{equation}
where $E_{\rm em,c}$ and $E_{\rm em,nc}$ are the EMF energy values for the collision and non-collision cases, respectively. The efficiency of the example case is shown in the lower panel of Figure \ref{fig:f1}, where we find that the efficiency is about 35\% near the end of collision process. This efficiency is much higher than the collision-induced kinetic energy release efficiency in the MFD outflows in the internal shock model of GRBs, which is typically a few percent or less \citep[e.g.][]{panaitescu99,kumar99,maxham09,GaoMeszaros15}. It is consistent with the analytic estimate of the ICMART model (\citealt{ZhangYan11}, see more discussion below in \S\ref{subsec:physical}). 

One important question is what mechanism causes this efficient magnetic energy dispassion? From the magnetic configuration we can see $B_x$ and $B_y$ have opposite directions around the collision region (see Figure \ref{fig:initial_cut}). We suggest that most likely the additional $E_{\rm em}$ dissipation is triggered by strong collision-driven reconnection events. In order to check our conjecture, in the following, we carry out a series of detailed analyses based on our simulation data.

The $E_{\rm em}$ evolution in Figure \ref{fig:f1} can be characterized in four stages: (1) an initial ``self adjustment" (steep decay) phase before $t\sim10$; (2) a ``plateau" phase from about $t\sim10$ to $t\sim38$; (3) a ``normal decay" phase from about $t\sim38$ to $t\sim120$; and (4) a final quasi-steady phase. We analyze these stages in detail below.

The major collision starts from the later part of the ``self-adjustment'' steep decay phase. The collision compresses the magnetic fields to make the energy level higher than non-collision case. Panel A of Figure \ref{fig:stages} shows a series of representative cuts at $t=4$. From left to right, the four images display the 3D current contour plot viewed from $Y$-axis, the 3D current contour plot viewed from $X$-axis, the 2D contour cut of the $y$-component of the outflow velocity ($V_y$) in the $YZ$-plane (x=0) corresponding to the current plot, and the 2D contour cut of the rest mass density in the $YZ$-plane (x=0), respectively. From these results we find that a strong current layer and a pair of outflows are forming around the contact surface, which are consistent with the features of a collision-driven reconnection. 

The second stage is the ``plateau" phase from about $t\sim10$ to $t\sim38$. Panel B of Figure \ref{fig:stages} shows a series of representative cuts at $t=18$. We can see that the current layer around the contact surface becomes clearer and more concentrated. The outflows become faster (nearly 0.75c) and are also more concentrated at the current layer. Besides the four representative cuts shown in all panels, for panel B, we also add one extra 2D contour cut of the $x$-component of the outflow velocity ($V_x$) in the $XZ$-plane (y=0) corresponding to the current plots, which presents another important result that the current layer actually generates multi-orientation outflows in a 3D structure. These results suggest that many mini-jets with relativistic speeds can be potentially generated, if multiple collisions are invoked in a PFD outflow. Another interesting phenomenon is that although the system undergoes a strong reconnection process which in principle dissipates the EMF energy significantly, the global $E_{\rm em}$ evolution is nearly flat and even shows slight increase during this stage. The main reason for this feature is that the initial strong reconnection is collision-driven. Besides the strong reconnection, collision-induced strong compression also exists and tends to increase $E_{\rm em}$, which balances and even slightly surpasses $E_{\rm em}$ dissipation due to reconnection. The additional outflow study in the following Section \S \ref{subsec:v_out}, which shows that the outflows become super-Alfv\'enic at this stage, also supports the above analysis.

The next stage is the ``normal decay" phase. We choose two series of representative cuts at $t=58$ (Panel C) and $t=94$ (Panel D), respectively. The current strength and outflow velocity are similar between panels C and D, while they are systematically weaker and slower compared with the ``plateau" phase (panel B). This means that the initial collision-driven effect becomes weaker and the reconnection-facilitated dissipation enters a relatively steady phase. In the mean time, compression becomes sub-dominant, so that globally $E_{\rm em}$ dissipates with a relatively steady rate, which roughly equals to $\frac{0.1E_{\rm em,0}}{40t_0}=\frac{c\cdot E_{\rm em,0}}{400L_0}$ in the center-of-mass frame of the blobs ($L_0$ is the length normalization factor introduced in Table \ref{tab:t2}). The additional outflow study in the following Section \S \ref{subsec:v_out}, which shows that the outflow velocity keeps being around the Alfv\'en velocity at this stage, also supports this conclusion.

Finally, after $t\sim120$, the reconnection-dissipation gradually becomes weaker, and the system enters the quasi-steady phase without obvious $E_{\rm em}$ dissipation. The $E_{\rm em}$ evolution becomes nearly parallel with the non-collision case in Figure \ref{fig:f1}.

From these analyses, we conclude that the collision between two high-$\sigma$ blobs can indeed trigger strong magnetic reconnections and dissipate a significant fraction of EMF energy due to the reconnection-facilitated processes.

\subsection{Additional outflow study}\label{subsec:v_out}

Following the above analyses, in this subsection, we carry out an additional study on the outflow velocity. We compare the local Lorentz factor of the outflow ($\Gamma_{\rm out}$) with the critical Lorentz factor $\Gamma_{\rm A}$ calculated from the local relativistic Alfv\'en velocity 
\begin{equation}
V_{\rm A} = \frac{c}{\sqrt{4 \pi h^{\prime}/B^{\prime 2}+1}},
\label{equ:va}
\end{equation}
and the critical Lorentz factor $\Gamma_{\rm ms}$ calculated from the maximum possible value of the local relativistic fast mode magnetosonic velocity 
\begin{equation}
V_{\rm ms} = \sqrt{V_{\rm A}^2 + C_s^2(1-V_{\rm A}^2/c^2)},
\label{equ:vms}
\end{equation}
where $h^{\prime}$ and $B^{\prime}$ are the specific enthalpy and magnetic strength in the local comoving frame of the fluid, and $C_s$ is the relativistic sound speed calculated by 
\begin{equation}
C_s=c\sqrt{\hat{\gamma}P/h^{\prime}}.
\end{equation} 

In order to investigate whether the fluid velocities exceed the two characteristic velocities, we define
\begin{eqnarray}
 R_{\rm A} & \equiv & \frac{\Gamma_{\rm out}}{\Gamma_{\rm A}},  \\
 R_{\rm ms} & \equiv & \frac{\Gamma_{\rm out}}{\Gamma_{\rm ms}}.
\end{eqnarray}

\begin{figure*}[!htb]
\plotone{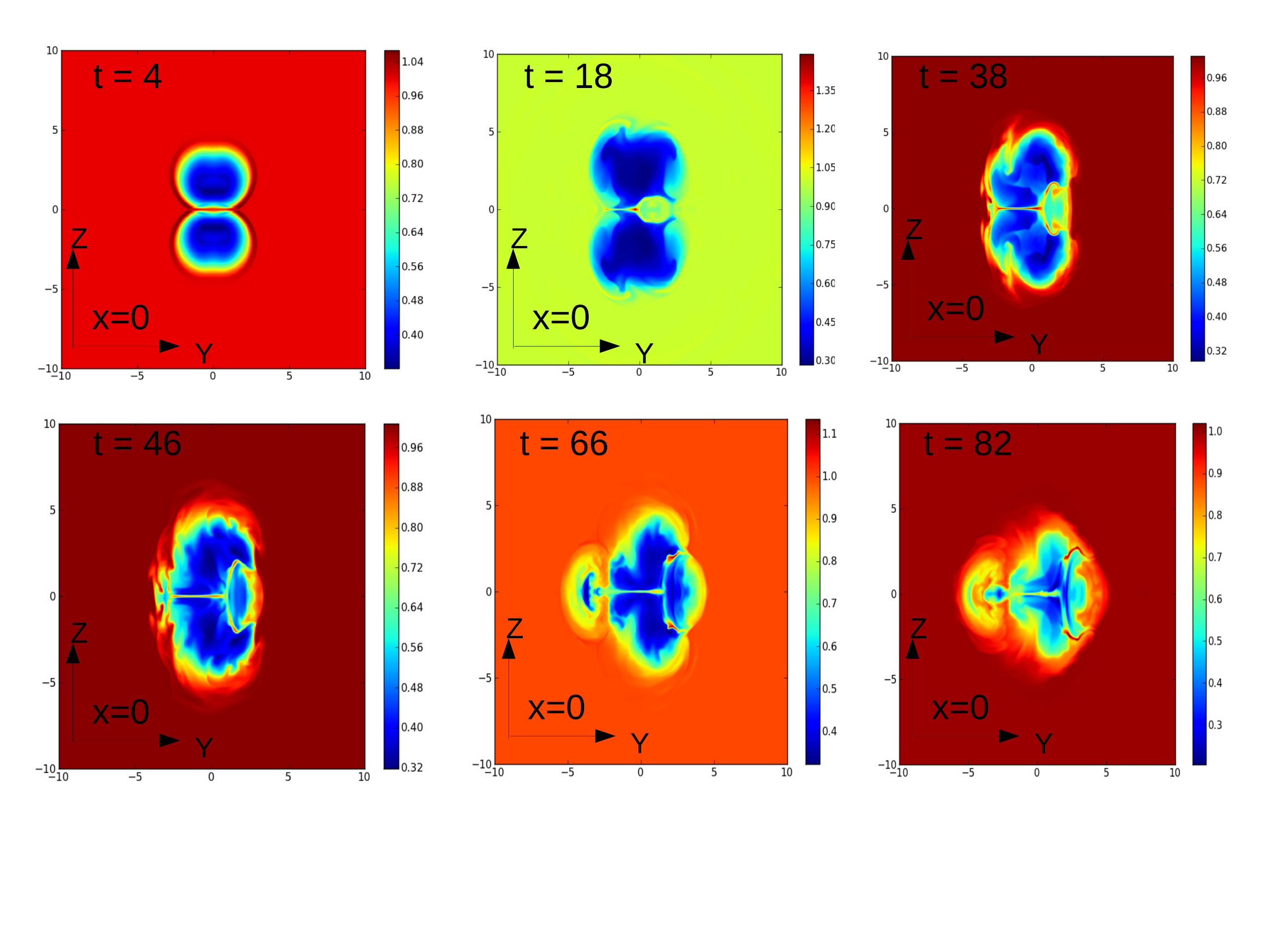}
\caption{The selected 2D contour cuts of $R_{\rm A}$ for different stages. The three panels in the upper row correspond to the starting time when $R_{\rm A}>1$, the time when $R_{\rm A}$ is the largest, and the ending time for the condition of $R_{\rm A}>1$, respectively. These correspond to the plateau stage. The three panels in the lower row correspond to three epochs during the normal decay phase, during which $\Gamma_{\rm out}$ becomes relatively steady and close to $\Gamma_{\rm A}$.}
\label{fig:ratio_va}
\end{figure*}

Figure \ref{fig:ratio_va} shows the selected 2D contour cuts of $R_{\rm A}$. The three panels in the upper row correspond to the starting time when $R_{\rm A}>1$ is reached ($t=4$), the time when $R_{\rm A}$ is the largest ($t=18$), and the ending time for the condition of $R_{\rm A}>1$ ($t=38$), respectively. After $t\sim38$, the $\Gamma_{\rm out}$ starts to become slightly smaller but still close to $\Gamma_{\rm A}$ (see the three panels in the lower row of Figure \ref{fig:ratio_va}). These results are consistent with the energy evolution analysis presented above in Section \S \ref{subsec:energy_evo}. The duration when $R_{\rm A}>1$ is satisfied is just the ``plateau" phase of energy evolution, in which strong compression exists and drives the outflows to become super-Alfv\'enic. After $t\sim 38$ the energy evolution enters the ``normal decay" phase, which corresponds to the phase of relatively steady reconnection-facilitated dissipation without strong compression, so that the outflow velocity is close to the theoretical Alfv\'enic velocity.

\begin{figure*}[!htb]
\plotone{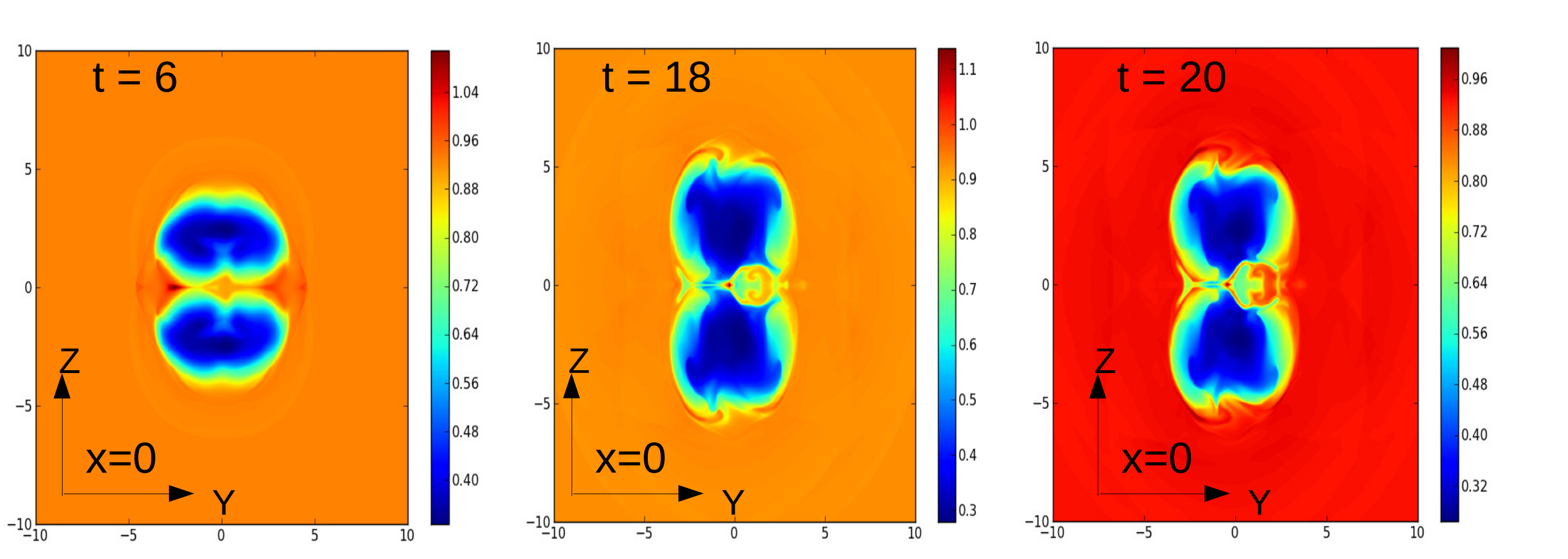}
\caption{the selected 2D contour cuts from the results of $R_{\rm ms}$. The three panels correspond to the starting time when $R_{\rm ms}>1$ is satisfied, the time when $R_{\rm ms}$ is the largest, and the ending time when the condition of $R_{\rm ms}>1$ is satisfied, respectively.}
\label{fig:ratio_vms}
\end{figure*}

Figure \ref{fig:ratio_vms} show the contour cuts of $R_{\rm ms}$. Since $V_{\rm ms}$ is the maximum wave propagation speed in a MHD system, if $R_{\rm ms}>1$, a local shock in the front of the outflow would potentially be generated. The three epochs shown in Figure \ref{fig:ratio_vms} correspond to the starting time when $R_{\rm ms}>1$ is satisfied ($t=6$), the time when $R_{\rm ms}$ is the largest ($t=18$), and the ending time for the condition of $R_{\rm ms}>1$ to be satisfied ($t=20$), respectively. These results indeed show a period of about 15 time units during which $R_{\rm ms}>1$ is satisfied. This duration is shorter than the duration when $R_{\rm A}>1$ is satisfied. For this case, the largest value of $R_{\rm ms}$ is about 1.13. Since the $\Gamma_{\rm out}$ depends on numerical resolution (see Section \S \ref{subsec:res} below for details) and other physical parameters, it is worthwhile to perform a more detailed study for this feature in the future. In this study, since $V_{\rm out}$ is only slightly larger than $V_{\rm ms}$ in a small local region and for a short duration, we do not resolve an obvious shock feature from the numerical data.

\subsection{Resolution study}\label{subsec:res}

\begin{figure}[!htb]
\plotone{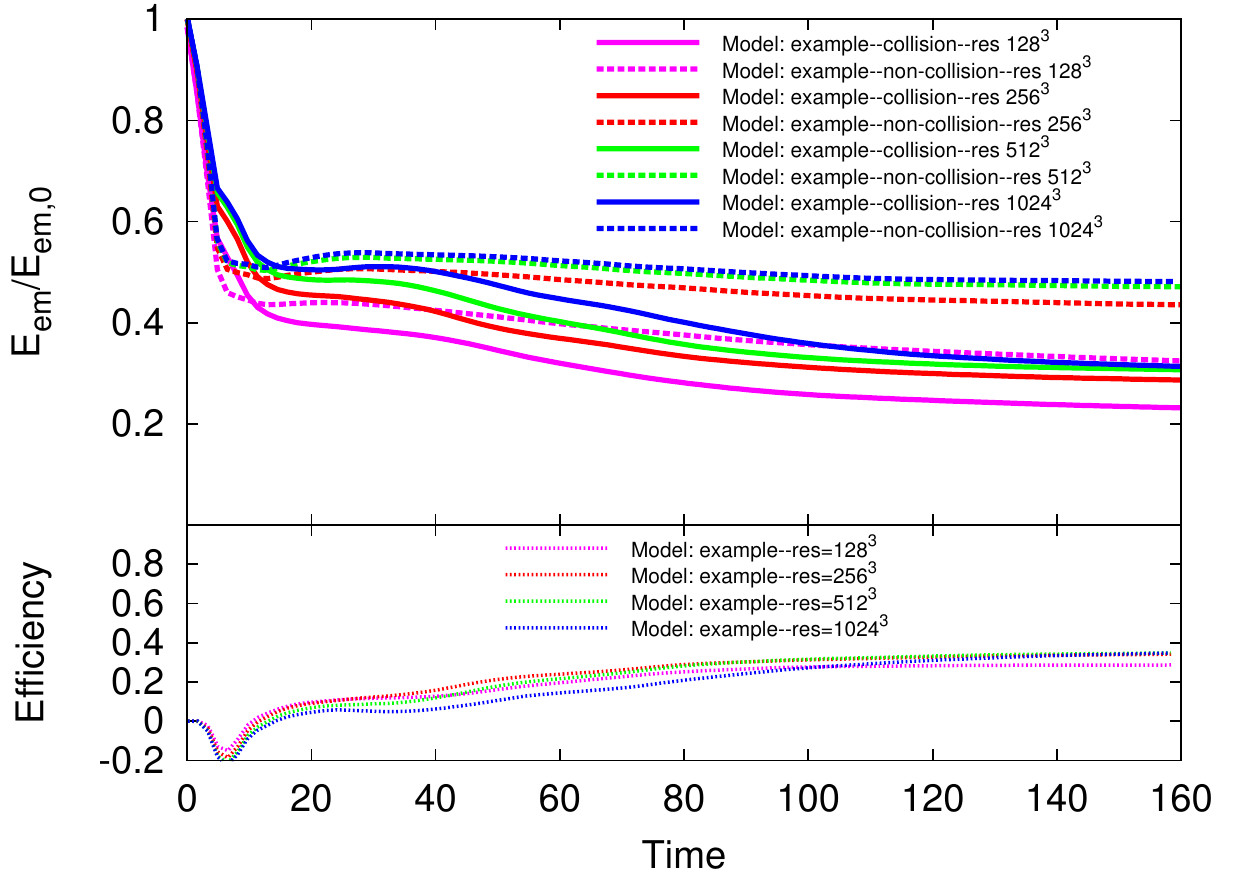}
\caption{A numerical resolution study based on the above example case in Section \S \ref{subsec:energy_evo}. The magenta, red, green and blue groups of lines correspond to the resolutions of $128^3$, $256^3$, $512^3$ and $1024^3$, respectively. The $E_{\rm em}$ dissipation efficiency at the finial quasi-steady phase is nearly the same in all cases.}
\label{fig:f_res}
\end{figure}

We now discuss the effects of numerical resolution on our results. Although the ideal MHD code that we use does not have explicit resistivity, it still has numerical resistivity from the numerical scheme, which depends on the resolution of the simulation. This may affect the reconnection rate and energy dissipation rate in the simulations. To address this uncertainty, we perform a resolution test based on the above example case. We keep the same box size and the parameters in Table \ref{tab:t1}, and only change the resolution. Figure \ref{fig:f_res} shows the results. The magenta, red, green and blue groups of lines correspond to the results with numerical resolution $128^3$, $256^3$, $512^3$ and $1024^3$, respectively. When the resolution decreases, we find that the level of $E_{\rm em}$ evolution is systematically lower and the efficiency also slightly decreases. On the other hand, the change of efficiency is only several percentage from the highest to the lowest resolutions, which means that the $E_{\rm em}$ dissipation efficiency is insensitive to numerical resolution. In addition, the $E_{\rm em}$ level and the efficiency in the final quasi-steady phase also show a trend of convergence when the resolution increases.

\begin{figure*}[!htb]
\plotone{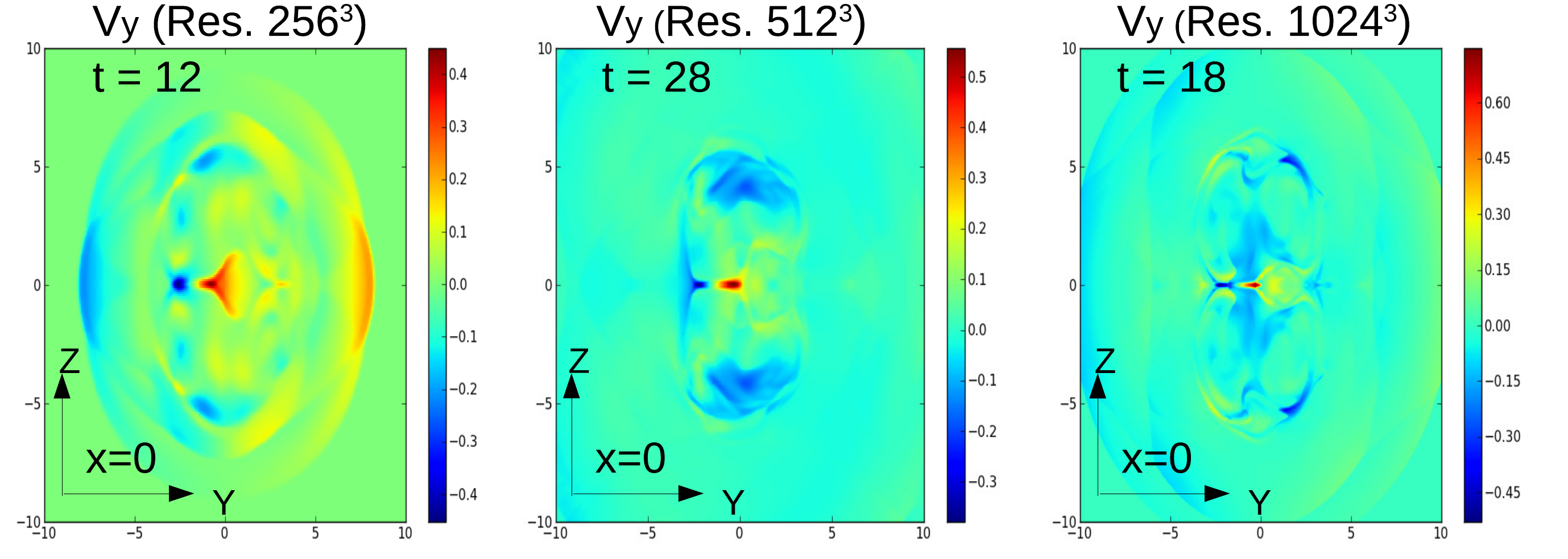}
\caption{The contour cuts corresponding to the maximum $y$-component of the outflow velocity ($V_y$) in the $YZ$-plane (x=0) for different resolutions. The maximum values of $V_y$ are about $0.45c$, $0.55c$, $0.75c$ for the resolutions of $256^3$, $512^3$, and $1024^3$, respectively. The aspect ratio becomes smaller for a higher resolution.}
\label{fig:f_max_vel}
\end{figure*}

Another important result from the resolution study is that the maximum outflow velocity increases when the resolution increases. Figure \ref{fig:f_max_vel} shows the contour cuts corresponding to the maximum $y$-component of the outflow velocity ($V_y$) in the $YZ$-plane (x=0) for different resolutions. The maximum values of $V_y$ are about $0.45c$, $0.55c$, and $0.75c$ for resolution of $256^3$, $512^3$, and $1024^3$, respectively. The reason is probably that the higher resolution decreases the effective numerical resistivity and decreases the aspect ratio between the thickness and the length of the reconnection layer, so that the outflow speed is forced to reach a higher value in order to balance the similar compression forced inflow. This analysis is also supported by Figure \ref{fig:f_max_vel}, which shows that with an increasing resolution, the length of the reconnection layer is similar, but the thickness becomes thinner.

\subsection{Physical analyses}\label{subsec:physical}

In this subsection we carry out some physical analyses to understand the $\sim$ 35\% $E_{\rm em}$ dissipation efficiency obtained from our numerical simulations.

Assuming a complete inelastic collision between two high-$\sigma$ blobs, \cite{ZhangYan11} analytically estimated the total efficiency of the collision-induced $E_{\rm em}$ dissipation efficiency based on energy and momentum conservation laws. Their Equation (51) can be written as
\begin{equation}
 \eta = \frac{1}{1+\sigma_{\rm b,f}} - \frac{\Gamma_m (m_1+m_2)}{(\Gamma_1 m_1 + \Gamma_2 m_2)(1+\sigma_{\rm b,i})},
\end{equation}
where $\sigma_{\rm b,i}$ is the initial $\sigma$ value of the two colliding blobs, $\sigma_{\rm b,f}$ is the final $\sigma$ value after the inelastic collision is over, $\Gamma_1$, $\Gamma_2$, and $\Gamma_m$ are the Lorentz factors of the two colliding blobs and the merged blob, respectively, and $m_1$, $m_2$ are the masses of the two colliding blobs. In our simulations, the two blobs are identical so that $m_1 = m_2$. Since we are observing in the merged frame so that $\Gamma_m = 1$, $\Gamma_1 = \Gamma_2 = \Gamma$, the final expression of the efficiency can be reduced to\footnote{This can be also derived directly by writing energy and momentum conservations in center-of-mass rest frame. }
\begin{equation}
 \eta = \frac{1}{1+\sigma_{\rm b,f}} - \frac{1}{\Gamma(1+\sigma_{\rm b,i})}.
\label{equ:eta2}
\end{equation}

\begin{figure*}[!htb]
\plotone{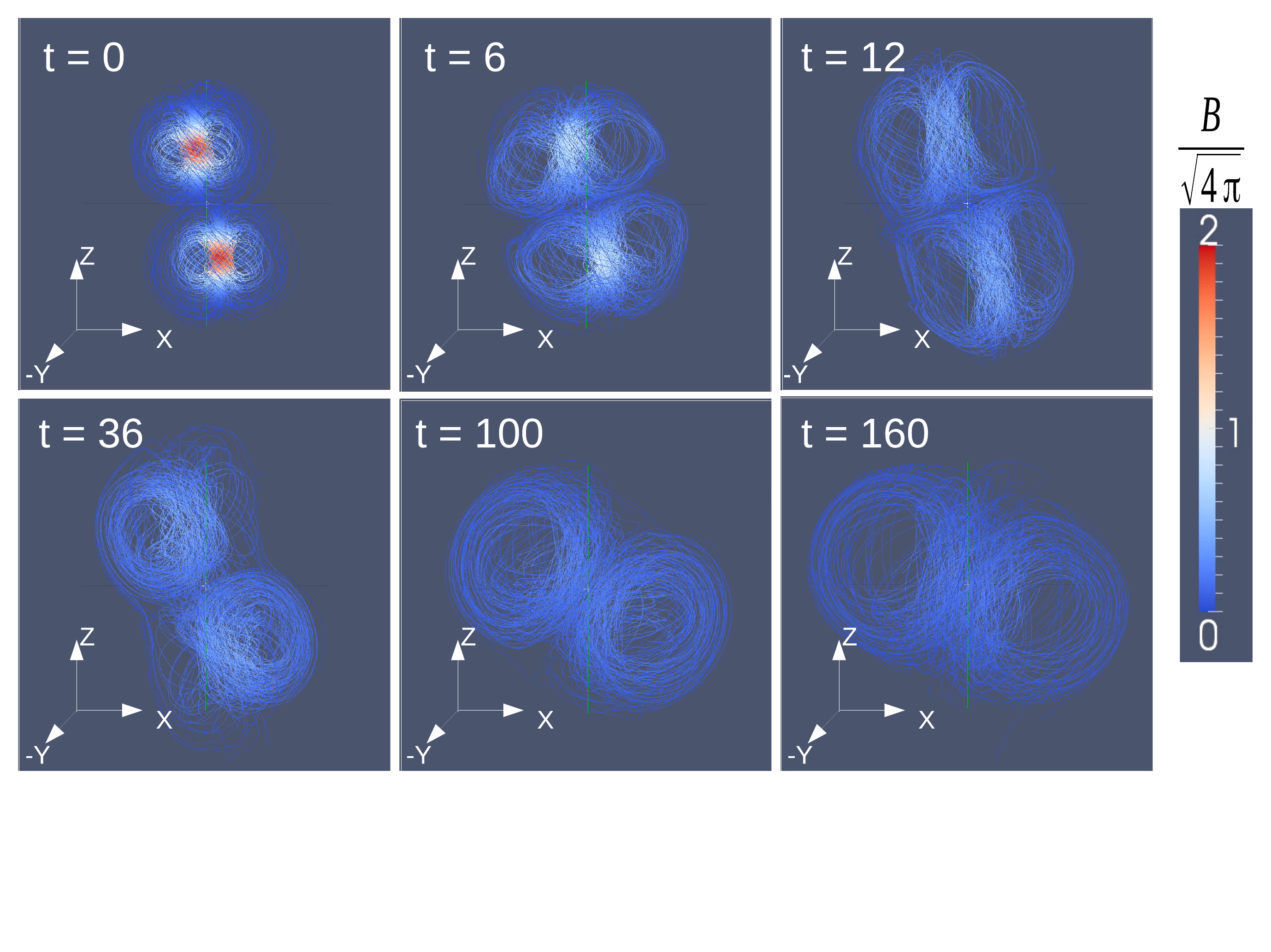}
\caption{The evolution of the field lines during the collision process. The two blobs merge into one larger blob, forming a ``$\infty$"-shaped field line configuration at the final quasi-steady stage of the evolution.}
\label{fig:field}
\end{figure*}

In order to connect this analytical equation with our simulated results, we first carry out some analyses to see if the condition of complete inelastic collision is satisfied. For ideal MHD simulations, fluid elements are attached to the field lines. Tracking the evolution of magnetic field configuration is therefore a convenient way to study whether collision is inelastic. Figure \ref{fig:field} shows several contour cuts of the 3D field line evolution. Initially the fields are compressed around $t=6$, and then bounce back around $t=12$. Later strong collision-driven reconnections on the contact surface efficiently dissipate the compressed magnetic energy and reduce the magnetic pressure in the center. This prevents further bouncing back and reorganizes the field configuration to make the two blobs merge into one larger blob with a new field configuration with a ``$\infty$" shape at the final quasi-steady stage. This suggests that the two blobs merge to one entity after the collision.

Due to the mis-alignment in $X$ direction of the two blobs, the collision would induce rotation during the merging process. This would render the collision process not completely inelastic. In order to investigate how important the rotation effect is, we calculate the ratio between the rotation energy ($E_{\rm rot}$) and the initial kinetic energy ($E_{\rm k,i}$). The rotation energy of the two blobs can be estimated as $E_{\rm rot}=2 \times (\frac{1}{2} I \omega^2)$, where the moment of inertia of one blob can be estimated as $I=\frac{2}{5} m r^2+m r^2$, where the first term denotes the moment of inertia of an idea sphere, and the second term denote the displacement from the rotation axis. Since the blobs expand with time, the size of the blob and its displacement increase with time. We estimate that after the merging process, $r$ is about three times of $r_0 = L_0 = 1$. We therefore derive $I \sim \frac{63}{5} m L_0^2=\frac{63}{5} m$. For the angular velocity $\omega$, we can estimate it from Figure \ref{fig:field}, which shows a roughly $\pi/4$ angular change within $\Delta t=90 L_0/c=90$. As a result, one can estimate $\omega\sim \pi/360$, so that $E_{\rm rot}=10^{-3}m$. The ratio between $E_{\rm rot}$ and $E_{\rm k,i}$ is therefore
\begin{equation}
\frac{E_{\rm rot}}{E_{\rm k,i}} = \frac{10^{-3}m}{2 \times \frac{1}{2} m V_{\rm b,z}^2} \approx 10^{-2}.
\label{equ:ratio_rot}
\end{equation}
So the rotation energy is only a small fraction of the initial kinetic energy, which means that the collision is very close to completely inelastic collision for this example case with $x_s=0.5$. While even if $E_{\rm rot}$ becomes a larger fraction of $E_{\rm k,i}$ when the misalignment $x_s$ increases, it would only reduce the kinetic energy dissipation efficiency, but would have little direct effect on the $E_{\rm em}$ dissipation efficiency that is our primary concern\footnote{However, the misalignment $x_s$ itself does have a direct effect on the $E_{\rm em}$ dissipation efficiency due to the different field configurations around the initial contact surface. See details in Section \ref{subsec:xs}.}. Due to the initial high-$\sigma$ property of the blobs, the contribution from the $E_{\rm k,i}$ dissipation to the total dissipation efficiency is only a minor fraction when $E_{\rm em}$ has significant dissipation, as we have found above.

With the above preparation, we can achieve a physical understanding of the high efficiency obtained from our simulation. Based on Eq.(\ref{equ:eta2}), we can derive the expected efficiency. From the initial condition, we derive $\Gamma=1.05$. From simulation results, we can also calculate $\sigma_{\rm b,f}$. Since $\sigma_{\rm b,f}$ has a complex spatial distribution, we perform a spatial average for all the positions with $\sigma_{\rm b,f}>1$ and also perform a time average from $t=90$ to $t=120$ to get the $\sigma_{\rm b,f} \approx 1.16$. As a result, we derive $\eta \approx 35.7\%$ based on the analytical calculation (Eq.(\ref{equ:eta2})). This is well consistent with the $E_{\rm em}$ dissipation efficiency calculated directly from the energy evolution of the simulations using Eq.(\ref{equ:phi}), as shown in Figure \ref{fig:f1}.

\subsection{Summary for this section}\label{subsec:summary_example}

In this section, we revealed a collision-induced strong reconnection process with the EMF energy dissipation efficiency about $35\%$, which is resolution insensitive. The outflow can locally become super-Alfv\'enic during the initial strong compression stage. The outflow velocity can potentially become relativistic in higher resolution simulations and generate multi-orientation mini-jets in a global PFD jet.

\section{Extended parameter space studies}\label{sec:para}

In Section \S \ref{sec:example}, we find significant EMF energy dissipation (about 35\%) facilitated by collision-driven magnetic reconnection. Based on the above analyses, we expect that some parameters may affect the results. First, the $\sigma_b$ evolution controls the $E_{\rm em}$ dissipation efficiency. The simulation results may then depend on the initial value $\sigma_{\rm b,i}$. Second, the initial misalignment $x_s$ gives different magnetic field configurations around the contact surface which may control the fraction of the free energy that can be released due to the reconnection processes. Next, different initial relative speed (kinetic energy) between the two blobs define the strengths of the initial collision-driven effect, so that it may be another factor to effect the conclusion. In addition, the initial displacement $z_d$ controls the delay of the collision. It is also interesting to investigate whether results depend on this parameter. Furthermore, the blobs undergo a significant expansion during the early ``self adjustment" phase before establishing a balance between the magnetic pressure force and the gas pressure force. Different background pressures and densities are therefore interesting input parameters that may affect the results. Finally, it is unknown whether the toroidal-to-poloidal ratio parameter, $\alpha$, plays a role to define the dissipation efficiency.

In this section, we perform a series of extended parameter studies to investigate the role of above-mentioned various parameters in defining the $E_{\rm em}$ dissipation efficiency and reconnection outflow properties. From the resolution study in Section \S \ref{subsec:res}, we find that the $E_{\rm em}$ dissipation efficiency is similar when the resolution is $\geqslant 256^3$. Since we are exploring a large parameter space, in order to reduce the simulation time, we use the $256^3$ resolution in all the simulations presented in this section. The general approach is that we only modify one parameter from the example simulation in each subsection, in order to explore the effects of that parameter. Below we explore the effect of following parameters in turn: the initial $\sigma$ value inside the blobs ($\sigma_{\rm b,i}$), the initial misalignment between the center of two blobs in $X$ direction ($x_s$), the initial velocity ($V_{\rm b,z}$) or relative Lorentz factor ($\Gamma_{\rm rel}$) of two blobs in $Z$ direction, the initial distance between the center of two blobs in $Z$ direction ($z_d$), the uniform gas pressure ($P$), the background density ($\rho_{\rm bkg}$), the toroidal-to-poloidal magnetic field ratio $\alpha$, and the adiabatic index ($\hat{\gamma}$).

\subsection{Initial $\sigma_{\rm b,i}$ of the blobs}\label{subsec:sigma}

\begin{figure}[!htb]
\plotone{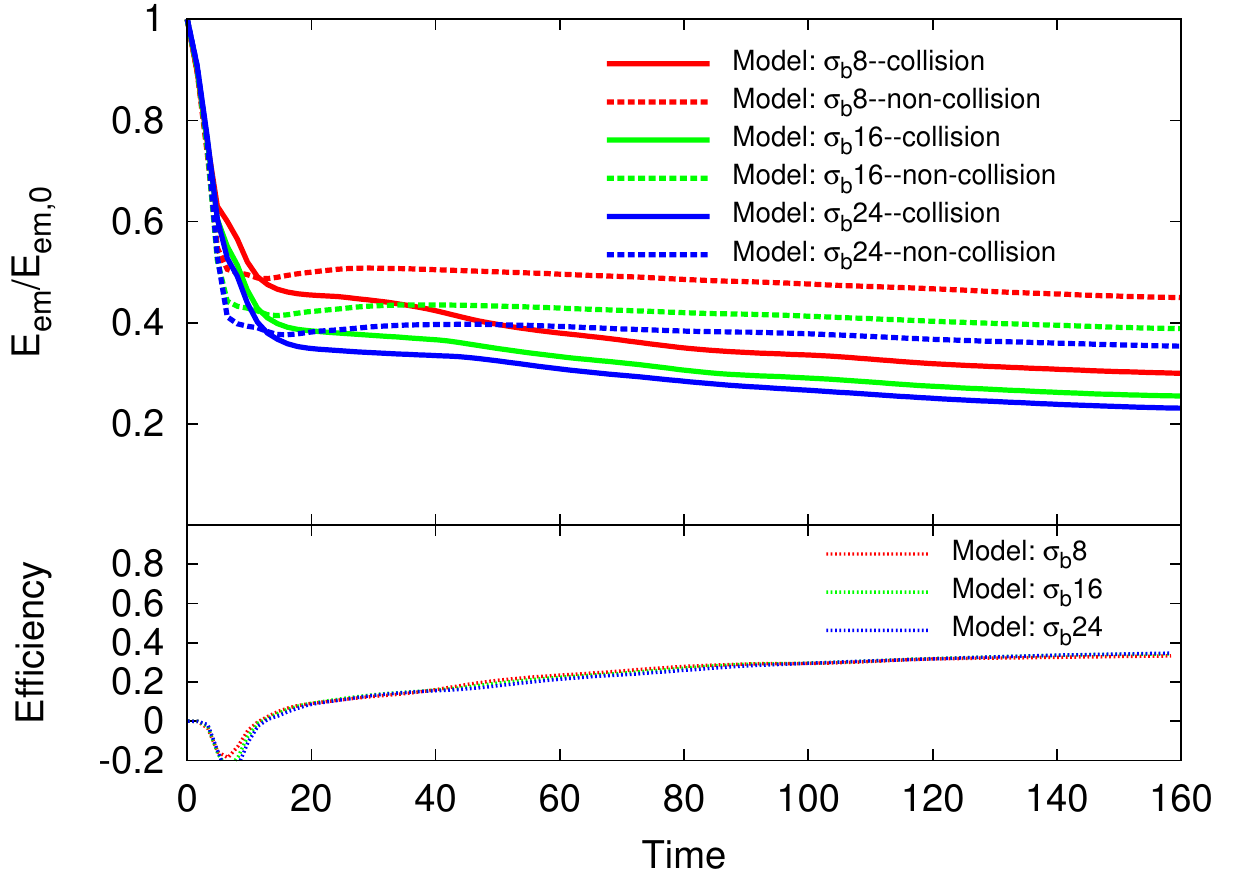}
\caption{The $E_{\rm em}$ evolution of three $\sigma_{\rm b,i}$ models: $\sigma_{\rm b,i}=8$ (red group), $\sigma_{\rm b,i}=16$ (green group), and $\sigma_{\rm b,i}=24$ (blue group). As the example model, we also show the non-collision cases corresponding to each of them to calculate the $E_{\rm em}$ dissipation efficiency. The $E_{\rm em}$ dissipation efficiency in the final quasi-steady phase is similar.}
\label{fig:sigma}
\end{figure}

The $\sigma_{\rm b,i}$ parameter affects the $\sigma_b$ evolution and the initial expansion of the blobs, so in this subsection we study the effect on the $E_{\rm em}$ energy evolution and dissipation efficiency for different $\sigma_{\rm b,i}$ values. We set a higher $\sigma_{\rm b,i}$ value by increasing the normalization parameter $B_{\rm b,0}$. The parameters we used are listed in Table \ref{tab:sigma}. Here the parameters besides $\sigma_{\rm b,i}$ and $B_{\rm b,0}$ are the same as the example model. We choose three different $\sigma_{\rm b,i}$ values. Here, due to the more significant expansion with a increasing $\sigma_{\rm b,i}$, we enlarge our simulation box to $30^3$ and also increase the resolution to $384^3$ to ensure the same absolute spatial resolution as the following sub-sections. The upper panel of Figure \ref{fig:sigma} shows the $E_{\rm em}$ evolution of these three models. Similar to the example model, we also show the non-collision case for each model to calculate the $E_{\rm em}$ dissipation efficiency. We find that with a higher $\sigma_{\rm b,i}$ value, the initial steep decay of $E_{\rm em}$ caused by the expansion is more significant due to the stronger outward magnetic pressure force. The $E_{\rm em}$ dissipation efficiency in the final quasi-steady phase, on the other hand, is rather similar for different $\sigma_{\rm b.i}$ values (lower panel of Figure \ref{fig:sigma}). In order to understand this result, we also calculate the $\sigma_{\rm b,f}$ values in the quasi-steady phase, and use Eq.(\ref{equ:eta2}) to calculate $\eta$ independently (see the method in Section \S \ref{subsec:physical}). The results are shown in Table \ref{tab:sigma_efficiency}. The calculated efficiencies have no obvious difference among different cases, since a larger $\sigma_{\rm b,i}$ corresponds to a slightly larger $\sigma_{\rm b,f}$, so that $\eta$ does not change significantly according to Eq.(\ref{equ:eta2}). This result is consistent with the efficiency calculated directly from the energy evolution of the simulations in the lower panel of Figure \ref{fig:sigma}.

\begin{table}[!htb]
\centering
\caption{The $\sigma_{\rm b,i}$-varying models}
\begin{tabular}{ccccccccc}
\toprule
Model name & $\sigma_{\rm b,i}$ & $B_{\rm b,0}$ & $\alpha$ & $\left|V_{\rm b,z}\right|$ & $P$ & $\rho_{\rm bkg}$ & $z_d$ & $x_s$ \\
\midrule
Model:$\sigma_{\rm b,i}$8 & 8 & $\sqrt{4\pi}$ & 3 & 0.3c & $10^{-2}$ & $10^{-1}$ & 4.4 & 1.0\\
Model:$\sigma_{\rm b,i}$16 & 16 & $\sqrt{8\pi}$ & 3 & 0.3c & $10^{-2}$ & $10^{-1}$ & 4.4 & 1.0\\
Model:$\sigma_{\rm b,i}$24 & 24 & $\sqrt{12\pi}$ & 3 & 0.3c & $10^{-2}$ & $10^{-1}$ & 4.4 & 1.0\\
\bottomrule
\end{tabular}
\label{tab:sigma}
\end{table}

\begin{table}[!htb]
\centering
\caption{$\sigma_{\rm b,f}$ - $\sigma_{\rm b,i}$ relation and the analytical vs. numerical efficiencies.}
\begin{tabular}{cccc}
\toprule
$\sigma_{\rm b,i}$ & $\sigma_{\rm b,f}$ & Efficiency & Efficiency\\
& & (analytical) & (numerical)\\
\midrule
8 & 1.16 & 35.7\% & 33.3\% \\
16 & 1.33 & 37.3\% & 34.4\% \\
24 & 1.49 & 36.4\% & 34.7\% \\
\bottomrule
\end{tabular}
\label{tab:sigma_efficiency}
\end{table}

\begin{figure}[!htb]
\plotone{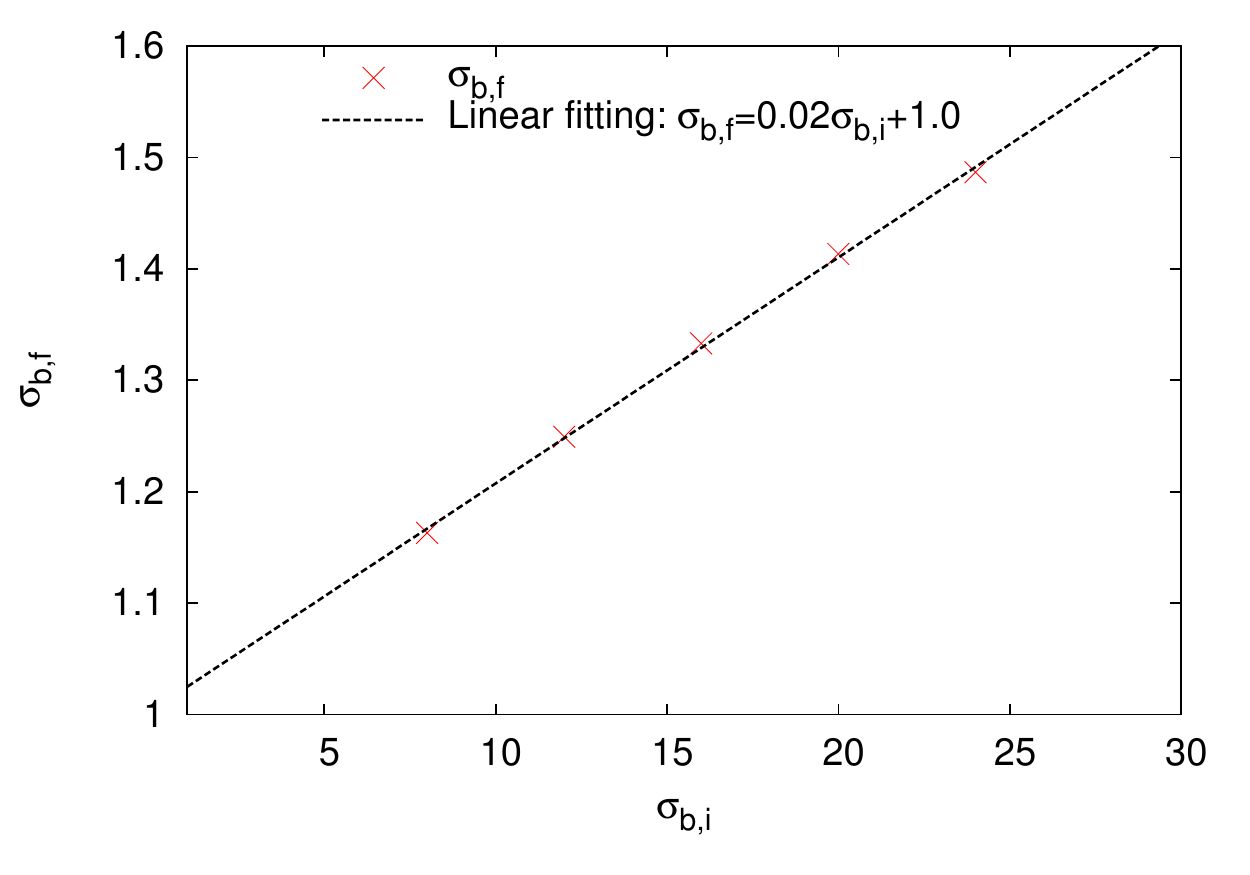}
\caption{The relationship between $\sigma_{\rm b,f}$ and $\sigma_{\rm b,i}$. The red cross points are the calculated results of $\sigma_{\rm b,f}$ corresponding to different $\sigma_{\rm b,i}$ from our simulations. The black dashed line is the linear fitting result.}
\label{fig:sigma_evo}
\end{figure}

One interesting result is that $\sigma_{\rm b,i}$ and $\sigma_{\rm b,f}$ values show a good linear relationship (Figure \ref{fig:sigma_evo}), which can be fitted by 
\begin{equation}
\sigma_{\rm b,f}=0.02\sigma_{\rm b,i}+1.0.
\label{equ:sigmas}
\end{equation} 
Physically, this equation does not apply for $\sigma_{\rm b,i}<1$. Right now the range of the $\sigma_{\rm b,i}$ is relatively small limited by the code capability. It is valuable to perform a more detailed study for a larger range of $\sigma_{\rm b,i}$ in the future to test this interesting and important relationship and to study the physical mechanism of this behavior.

\subsection{Initial misalignment between two blobs ($x_s$)}\label{subsec:xs}

The initial misalignment affects the magnetic field configuration around the contact surface during the collision and also the rotation property, so in this subsection we study the effect of $x_s$ on the $E_{\rm em}$ energy evolution and dissipation efficiency. The parameters we used are listed in Table \ref{tab:xs}. Besides $x_s$, other parameters are the same as the example model. From the analyses in Section \S \ref{subsec:physical}, the typical radius of one blob after expansion is $r\approx 3 r_0=3.0$. In order to make a relatively significant collision, $x_s$ should be smaller than $2r \approx 6$. We choose seven different values of $x_s$ in the simulations (Table \ref{tab:xs}). Among these models, ``Model:$x_s$1.0" is the same as the example model with resolution $256^3$.

\begin{figure}[!htb]
\plotone{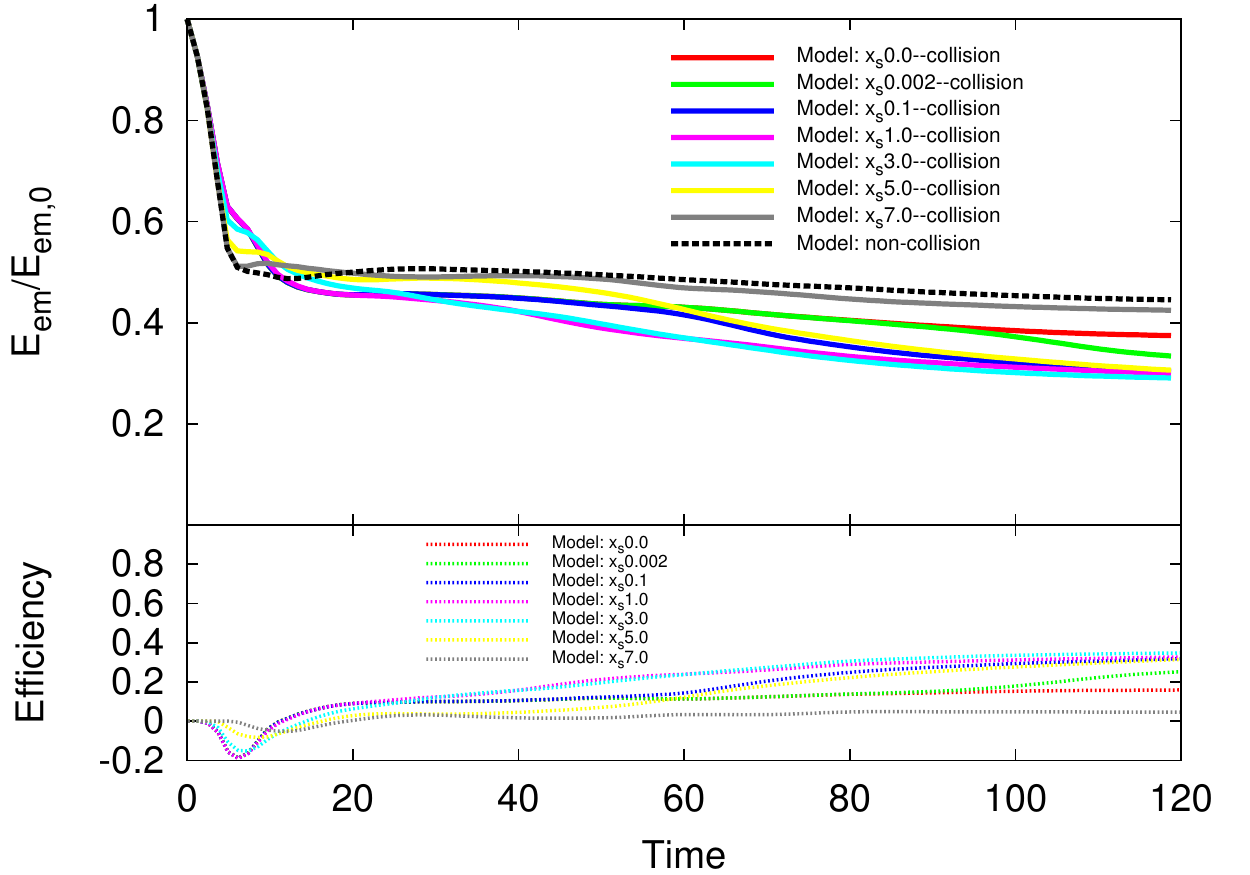}
\caption{The $E_{\rm em}$ evolution of six $x_s$ models: $x_s=0$ (red), $x_s=0.002$ (green), $x_s=0.1$ (blue), $x_s=1.0$ (magenta), $x_s=3.0$ (cyan), $x_s=5.0$ (yellow), and $x_s=7.0$ (gray). Similar to the example model, we also show the non-collision cases for each model to calculate the $E_{\rm em}$ dissipation efficiency. Since the non-collision case does not depend on $x_s$, there is only one non-collision evolution case (dashed line). Besides the $x_s$7.0 model which no collision is observed, Only the ``$x_s$0.0" model has obvious difference in $E_{\rm em}$ dissipation efficiency, which is only about half of the other models in the final quasi-steady phase.}
\label{fig:xs}
\end{figure}

The upper panel of Figure \ref{fig:xs} shows the $E_{\rm em}$ evolution of these seven models. We find that, with increasing $x_s$, the $E_{\rm em}$ dissipation efficiency first increases, and then decreases after reaching the maximum efficiency around $x_s \approx 3.0$. Such a behavior can be understood using the two lower panels in Figure \ref{fig:initial_cut}. The 2D cuts of $B_x$, $B_y$ show the directions and strength distributions of these two components of the magnetic field. Due to the initial expansion, the actual size of these configuration during collision would become about three times larger than the initial cuts. We consider the anti-parallel regions as the free energy source for reconnection-induced dissipation. When $x_s=0$, only the $B_x$ component can be reconnected. With an increasing $x_s$ from 0 to 3.0, the anti-parallel region of $B_x$ becomes smaller while the anti-parallel region of $B_y$ increases. Since the strength of $B_y$ is larger than $B_x$, the total dissipated magnetic energy becomes larger, which increases the dissipation efficiency to a higher value. When $x_s=3.0$, the anti-parallel region of $B_y$ reaches the maximum so that the maximum efficiency is achieved. After this critical point, the anti-parallel region of $B_y$ decreases with increasing $x_s$, which leads to a decrease in efficiency. Finally, when $x_s$ becomes larger than the size of the two blobs (6.0), the $E_{\rm em}$ evolution is nearly the same as the non-collision case due to the lack of significant collision between the two blobs.

\begin{figure*}[!htb]
\plotone{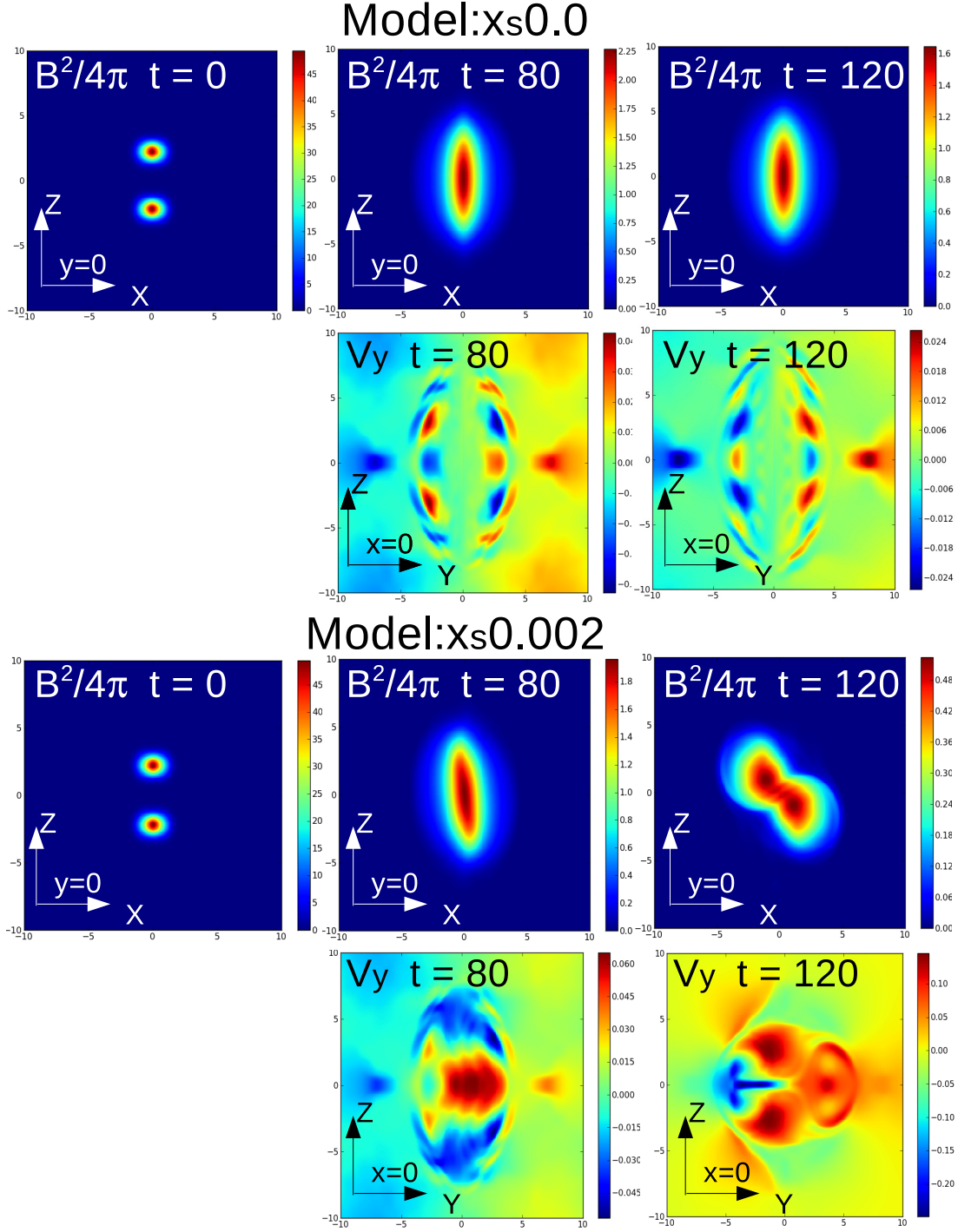}
\caption{Selected cuts for model ``Model:$x_s$0.0" (first two rows) and ``Model:$x_s$0.002" (last two rows). For each model, we draw the 2D contour cuts of $B^2/4\pi$ in the $XZ$-plane (y=0) at $t=0$, $t=80$ and $t=120$ (higher row) and the 2D contour cuts of the $y$-component of outflow velocity ($V_y$) in the $YZ$-plane (x=0) at $t=80$ and $t=120$ (lower row).  The results show that even if there is a very small misalignment, the merged blobs would start to rotate at some critical point and trigger additional reconnection-facilitated magnetic dissipation.}
\label{fig:xs_cuts}
\end{figure*}

Another important feature is that the change of efficiency as a function of $x_s$ is not linear. In fact, in our studied the cases only the two extreme cases $x_s=0$ and $x_s=7.0$ have significantly lower efficiencies compared with the other cases. Even if there is a very small misalignment, e.g. $x_s=0.002$, the efficiency could become significantly larger than the $x_s=0$ case. Inspecting the energy evolution plot (Fig.\ref{fig:xs}), one can see that the main difference comes from the ``normal decay" phase. A smaller $x_s$ would delay the ``normal decay" phase to a later time, whereas there is no ``normal decay" phase if $x_s=0$. In order to understand this feature, we draw Figure \ref{fig:xs_cuts}. The three panels in the first row are the 2D contour cuts of $B^2/4\pi$ in the $XZ$-plane (y=0), corresponding to $t=0$, $t=80$, and $t=120$ for ``Model:$x_s$0.0", respectively. The following two panels in the second row are the 2D contour cuts of the $y$-component of the outflow velocity ($V_y$) in the $YZ$-plane (x=0), corresponding to $t=80$ and $t=120$ for ``Model:$x_s$0.0". The following five panels are the corresponding cuts for the ``Model:$x_s$0.002". One can see that for the complete alignment case (``Model:$x_s$0.0"), the two blobs merge together and enter the quasi-steady phase without an obvious reconnection feature. For ``Model:$x_s$0.002", on the other hand, the merged blobs start to rotate at $t \sim 80$, which just corresponds to the starting time in the energy evolution plot (Fig.\ref{fig:xs}) when the two models become distinctly different.  The last panel of $V_y$ in Figure \ref{fig:xs_cuts} shows the feature of the reconnection-driven outflows during the ``normal decay" phase of ``Model:$x_s$0.002", which is not seen in ``Model:$x_s$0.0\footnote{The outflows look like asymmetric on that 2D cut, but the 3D configuration is more complex.}. From these analyses, we can draw conclusion that, besides the initial anti-parallel region caused by $x_s$, the collision-triggered rotation is another important process to change the magnetic field configuration and dissipate more magnetic energy. This rotation-driven dissipation seems to correspond to the ``normal decay" phase in the energy evolution plot.

\begin{table}[!htb]
\centering
\caption{The $x_s$-varying models}
\begin{tabular}{ccccccccc}
\toprule
Model name & $\sigma_{\rm b,i}$ & $B_{\rm b,0}$ & $\alpha$ & $\left|V_{\rm b,z}\right|$ & $P$ & $\rho_{\rm bkg}$ & $z_d$ & $x_s$ \\
\midrule
Model:$x_s$0.0 & 8 & $\sqrt{4\pi}$ & 3 & 0.3c & $10^{-2}$ & $10^{-1}$ & 4.4 & 0\\
Model:$x_s$0.002 & 8 & $\sqrt{4\pi}$ & 3 & 0.3c & $10^{-2}$ & $10^{-1}$ & 4.4 & 0.002\\
Model:$x_s$0.02 & 8 & $\sqrt{4\pi}$ & 3 & 0.3c & $10^{-2}$ & $10^{-1}$ & 4.4 & 0.1\\
Model:$x_s$0.1 & 8 & $\sqrt{4\pi}$ & 3 & 0.3c & $10^{-2}$ & $10^{-1}$ & 4.4 & 1.0\\
Model:$x_s$1.0 & 8 & $\sqrt{4\pi}$ & 3 & 0.3c & $10^{-2}$ & $10^{-1}$ & 4.4 & 3.0\\
Model:$x_s$1.6 & 8 & $\sqrt{4\pi}$ & 3 & 0.3c & $10^{-2}$ & $10^{-1}$ & 4.4 & 5.0\\
Model:$x_s$1.6 & 8 & $\sqrt{4\pi}$ & 3 & 0.3c & $10^{-2}$ & $10^{-1}$ & 4.4 & 7.0\\
\bottomrule
\end{tabular}
\label{tab:xs}
\end{table}

\subsection{Initial relative Lorentz factor ($\Gamma_{\rm rel}$) between the two blobs}\label{subsec:vel}

A larger initial velocity $V_{\rm b,z}$ means a larger initial kinetic energy of the two blobs, which would provide a larger driving force initially and also effectively decrease $\sigma_{\rm b,i}$. In this subsection, we study the effect of $V_{\rm b,z}$ in detail. Since we are testing $V_{\rm b,z}$ in the relativistic regime, we adopt the relative Lorentz factor ($\Gamma_{\rm rel}$) between two blobs as the varying parameter. The relative Lorentz factor can be calculated as
\begin{equation}
\Gamma_{\rm rel}=2\Gamma_{\rm b,z}^2-1,
\end{equation} 
where $\Gamma_{\rm b,z} = ({1-V_{\rm b,z}^2/c^2})^{-1/2}$ is the Lorentz factor of each blob.

\begin{figure}[!htb]
\plotone{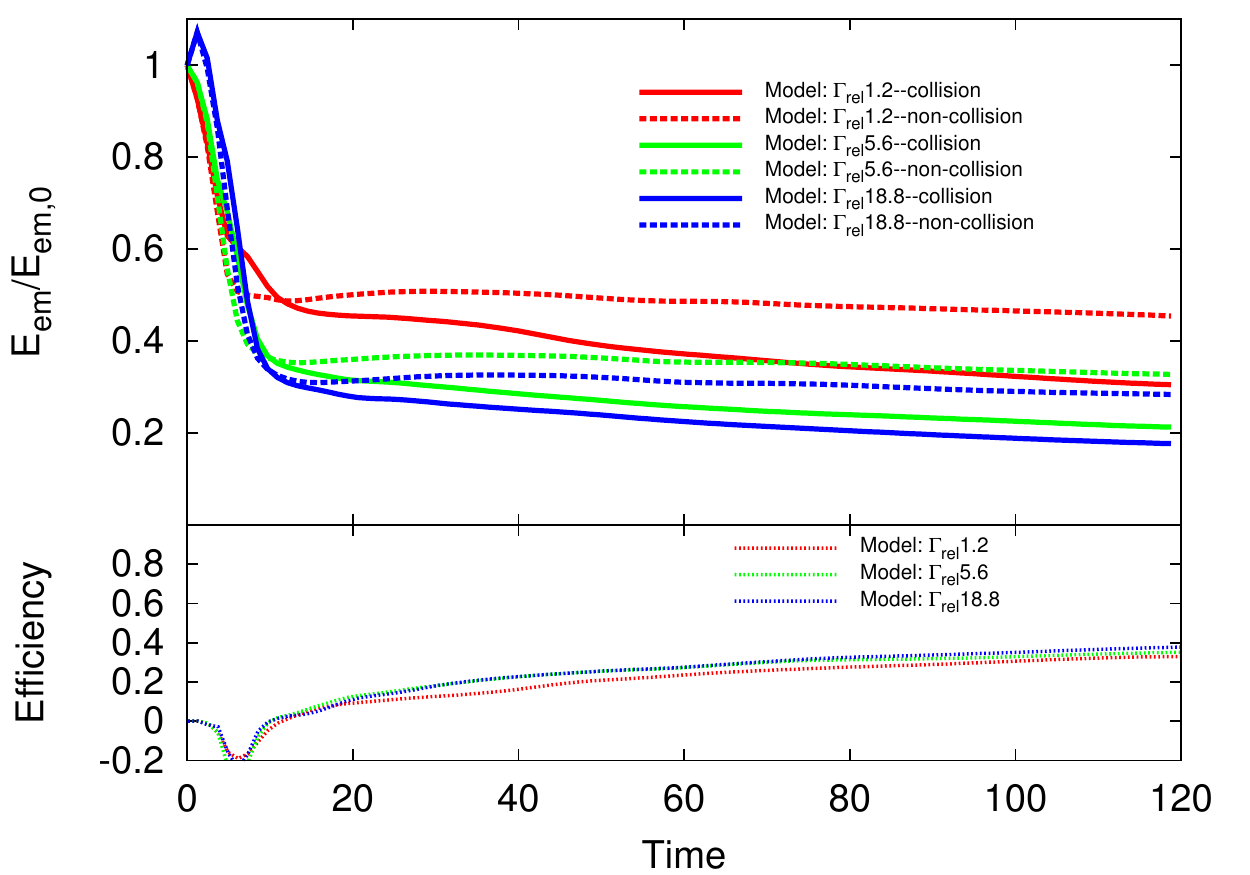}
\caption{The $E_{\rm em}$ evolution of four $\Gamma_{\rm rel}$ models: $\Gamma_{\rm rel}=1.2$ (red), $\Gamma_{\rm rel}=5.6$ (green), and $\Gamma_{\rm rel}=18.8$ (blue). Similar to the example model, for each model, we also show the non-collision case (dashed) to calculate the $E_{\rm em}$ dissipation efficiency. the $E_{\rm em}$ dissipation efficiency 
is larger for the model with a larger $\Gamma_{\rm rel}$.}
\label{fig:vel}
\end{figure}

The parameters we used are listed in Table \ref{tab:vel}. The parameters except $\Gamma_{\rm rel}$ and $\sigma_{\rm b,i}$ are the same as the example model. The effective change of $\sigma_{\rm b,i}$ is a consequence of changing $\Gamma_{\rm rel}$, since we keep $B$ as observed in the center-of-mass frame as constant. We choose three different $\Gamma_{\rm rel}$ values. With an increasing $\Gamma_{\rm rel}$, the fraction of the initial kinetic energy becomes larger, so that $\sigma_{\rm b,i}$ effectively decreases. The upper panel of Figure \ref{fig:vel} shows the $E_{\rm em}$ evolution of these three models. As $\Gamma_{\rm rel}$ increases, due to the stronger dynamic process, the initial ``self adjustment" phase becomes more significant. In particular, for $\Gamma_{\rm rel}=18.8$, the fast motion generates a strong shock into the surrounding medium. The shock compresses the magnetic field even further, so that $E_{\rm em}$ reaches an even higher value initially. In addition, from the lower panel of Figure \ref{fig:vel}, we find that the $E_{\rm em}$ dissipation efficiency in the final quasi-steady phase increases with increasing $\Gamma_{\rm rel}$. This is because a higher initial $\Gamma_{\rm rel}$ carries a larger kinetic energy and gives a stronger initial collision-induced compression and reconnection-induced dissipation.

\begin{table}[!htb]
\centering
\caption{The $\Gamma_{\rm rel}$-varying models}
\begin{tabular}{ccccccccc}
\toprule
Model name & $\sigma_{\rm b,i}$ & $B_{\rm b,0}$ & $\alpha$ & $\Gamma_{\rm rel}$ & $P$ & $\rho_{\rm bkg}$ & $z_d$ & $x_s$ \\
\midrule
Model:$\Gamma_{rel}$1.2 & 9 & $\sqrt{4\pi}$ & 3 & 1.2 & $10^{-2}$ & $10^{-1}$ & 4.4 & 1.0\\
Model:$\Gamma_{rel}$5.6 & 6 & $\sqrt{4\pi}$ & 3 & 5.6 & $10^{-2}$ & $10^{-1}$ & 4.4 & 1.0\\
Model:$\Gamma_{rel}$18.8 & 3 & $\sqrt{4\pi}$ & 3 & 18.8 & $10^{-2}$ & $10^{-1}$ & 4.4 & 1.0\\
\bottomrule
\end{tabular}
\label{tab:vel}
\end{table}

\subsection{Initial distance between two blobs ($z_d$)}\label{subsec:zd}

\begin{figure}[!htb]
\plotone{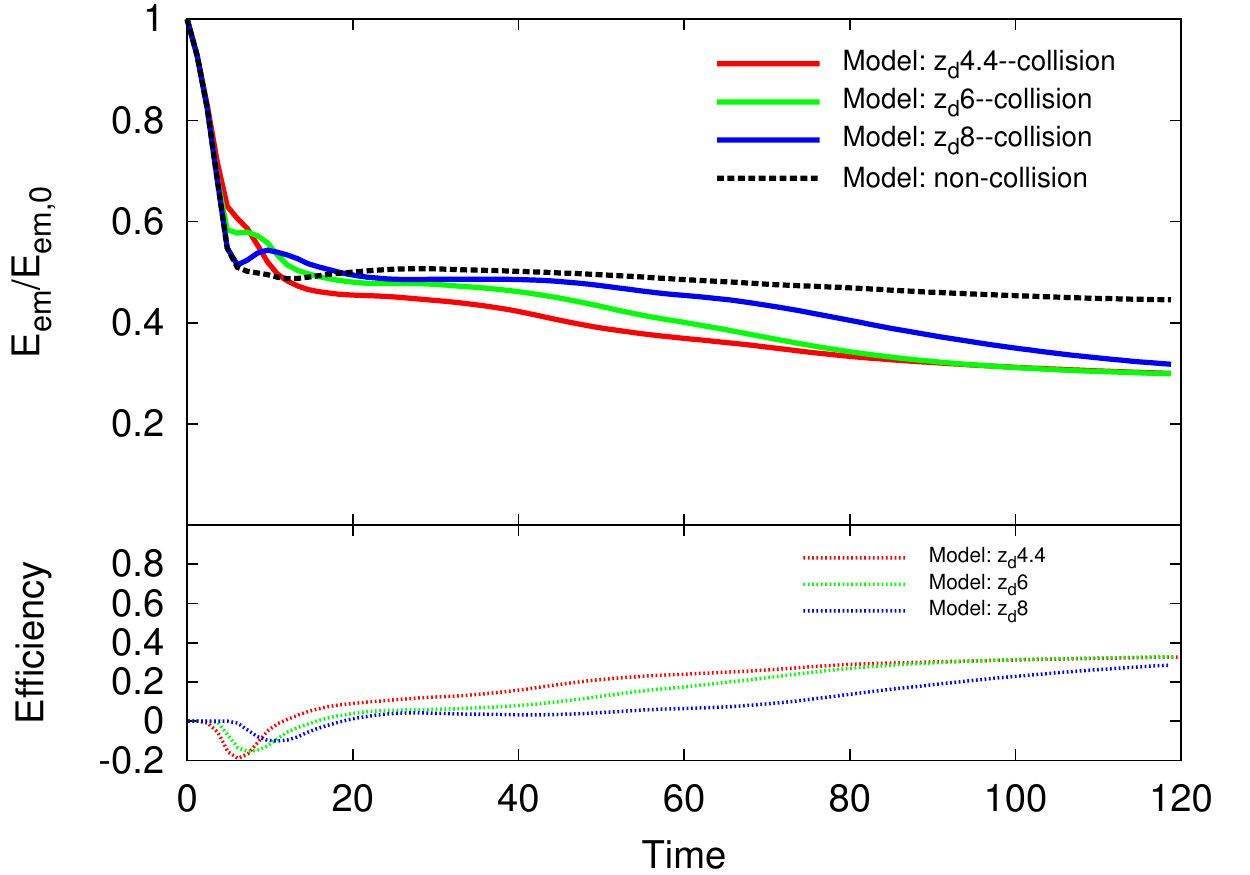}
\caption{The $E_{\rm em}$ evolution of three $z_d$ models: $z_d=4.4$ (red), $z_d=6$ (green), and $z_d=8$ (blue), which correspond to the collision happening at different stage of the blob evolution. Similar to the example model, the non-collision case (black dashed, the same for all three models) is plotted for comparison. Although the reconnection facilitated dissipation is delayed when $z_d$ increases, the energy level and the $E_{\rm em}$ dissipation efficiency in the final quasi-steady phase are similar.}
\label{fig:zd}
\end{figure}

The initial distance between the center of two blobs in z direction ($z_d$) controls the delay of the collision. In our example case, $z_d$ is relatively small and the collision happens around the middle stage of the initial expansion. It is valuable to study the effect on the $E_{\rm em}$ energy evolution and dissipation efficiency when we set a larger $z_d$ to delay the collision time to a later stage. The parameters we used are listed in Table \ref{tab:zd}. Here the parameters except $z_d$ are the same as the example model. We choose three different $z_d$ values. The ``Model:$z_d$4.4" uses $z_d=4.4$ which is the same as the example model with resolution $256^3$, and the other two models, ``Model:$z_d$6" and ``Model:$z_d$8", have $z_d=6,8$, respectively. The upper panel of Figure \ref{fig:zd} shows the $E_{\rm em}$ evolution of these three models. The collision times of these three different $z_d$ models are designated to be around the middle and late stages of initial expansion as well as after the initial expansion, respectively. We find that the reconnection-triggered $E_{\rm em}$ dissipation process is systematically delayed when $z_d$ becomes larger. However, the efficiency at the final quasi-steady phase reaches a similar value (lower panel of Figure \ref{fig:zd}) in all three models. This result suggests that the dissipation efficiency does not sensitively depend on the stage of blob evolution.

\begin{table}[!htb]
\centering
\caption{The $z_d$-varying models}
\begin{tabular}{ccccccccc}
\toprule
Model name & $\sigma_{\rm b,i}$ & $B_{\rm b,0}$ & $\alpha$ & $\left|V_{\rm b,z}\right|$ & $P$ & $\rho_{\rm bkg}$ & $z_d$ & $x_s$ \\
\midrule
Model:$z_d$4.4 & 8 & $\sqrt{4\pi}$ & 3 & 0.3c & $10^{-2}$ & $10^{-1}$ & 4.4 & 1.0\\
Model:$z_d$6 & 8 & $\sqrt{4\pi}$ & 3 & 0.3c & $10^{-2}$ & $10^{-1}$ & 6 & 1.0\\
Model:$z_d$8 & 8 & $\sqrt{4\pi}$ & 3 & 0.3c & $10^{-2}$ & $10^{-1}$ & 8 & 1.0\\
\bottomrule
\end{tabular}
\label{tab:zd}
\end{table}

\subsection{Background gas pressure ($P$)}\label{subsec:pre}

\begin{figure}[!htb]
\plotone{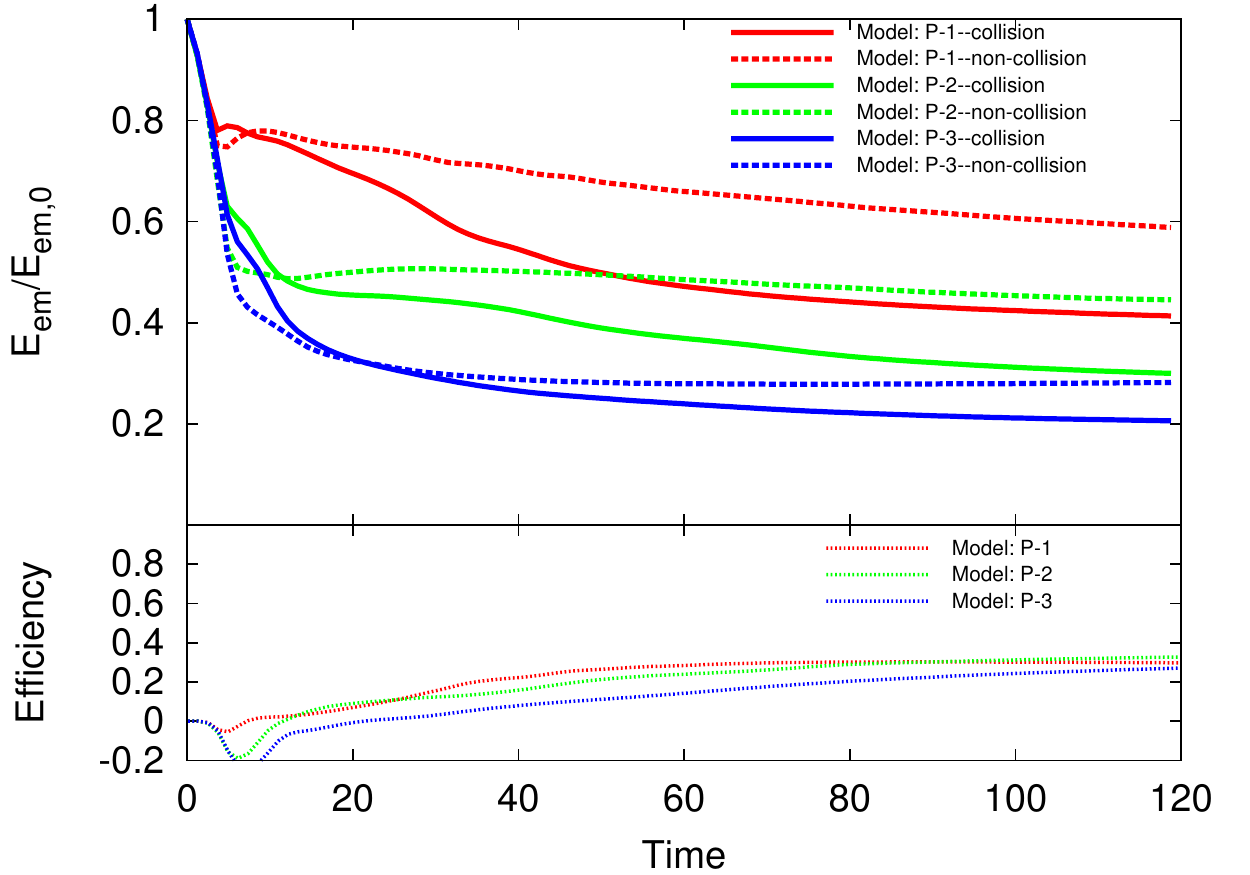}
\caption{The $E_{\rm em}$ evolution of three pressure models: $P=10^{-1}$ (red), $P=10^{-2}$ (green) and $P=10^{-3}$ (blue). Similar to the example model, we also show the non-collision cases (dashed lines) corresponding to each pressure model to calculate the $E_{\rm em}$ dissipation efficiency. Although the initial ``self adjustment" phase due to the expansion becomes more significant when the pressure goes to lower values, the difference of efficiency in the final quasi-steady phase is relative small.}
\label{fig:pressure}
\end{figure}

The blobs undergo a significant initial expansion to establish a balance between the magnetic pressure and the ambient gas pressure. Different background pressure ($P$) and density ($\rho_{\rm bkg}$) would affect these processes. In this and next subsections, we study the effect of $P$ and $\rho_{\rm bkg}$ on the $E_{\rm em}$ energy evolution and dissipation efficiency. The parameters for the $P$-varying models are listed in Table \ref{tab:pressure}. Here the parameters other than $P$ are the same as the example model. We choose three pressure values. The ``Model:P-2" is the same as the example model with resolution $256^3$, and the other two models have a lower (Model:P-1) or higher (Model:P-3) pressure. The upper panel of Figure \ref{fig:pressure} shows the $E_{\rm em}$ evolution of these three models. The initial ``self adjustment" phase due to the expansion is more significant when pressure becomes lower, since for a lower background pressure it takes longer for the magnetic blob to expand before reaching a balance with the ambient gas, and vice versa. Even with very different blob dynamics for different pressure values, the difference of $E_{\rm em}$ dissipation efficiency in the final quasi-steady phase is not so large as the $E_{\rm em}$ evolution itself (lower panel of Figure \ref{fig:pressure}), which means that the efficiency is relatively insensitive to the expansion process. This is probably due to the fact that the initial expansion phase with different background pressure values affects the evolution of both the collision and non-collision cases to similar degrees, so that the relative difference (efficiency) does not have a significant change.

\begin{table}[!htb]
\centering
\caption{The $P$-varying models}
\begin{tabular}{ccccccccc}
\toprule
Model name & $\sigma_{\rm b,i}$ & $B_{\rm b,0}$ & $\alpha$ & $\left|V_{\rm b,z}\right|$ & $P$ & $\rho_{\rm bkg}$ & $z_d$ & $x_s$ \\
\midrule
Model:P-1 & 8 & $\sqrt{4\pi}$ & 3 & 0.3c & $10^{-1}$ & $10^{-1}$ & 4.4 & 1.0\\
Model:P-2 & 8 & $\sqrt{4\pi}$ & 3 & 0.3c & $10^{-2}$ & $10^{-1}$ & 4.4 & 1.0\\
Model:P-3 & 8 & $\sqrt{4\pi}$ & 3 & 0.3c & $10^{-3}$ & $10^{-1}$ & 4.4 & 1.0\\
\bottomrule
\end{tabular}
\label{tab:pressure}
\end{table}

\subsection{Background density ($\rho_{\rm bkg}$)}\label{subsec:den}

\begin{figure}[!htb]
\plotone{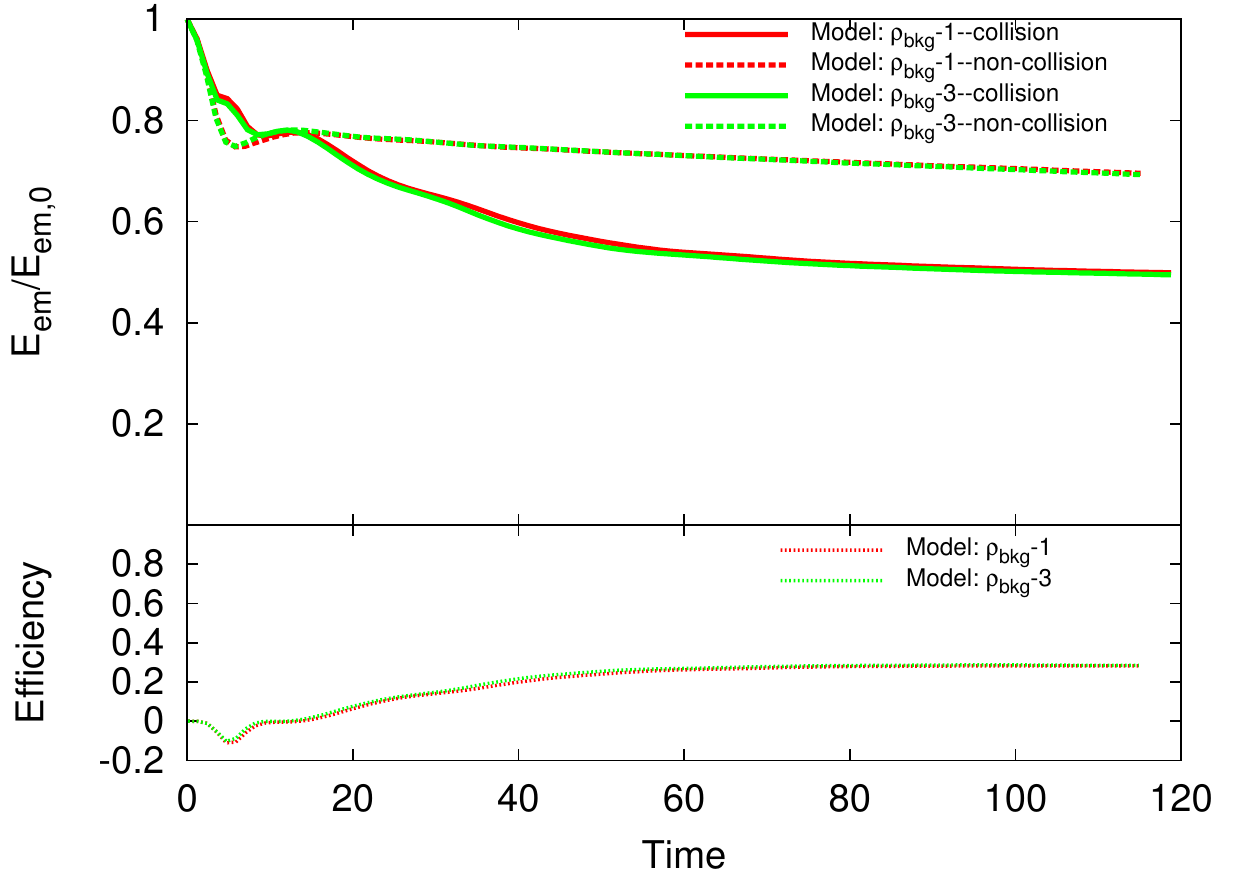}
\caption{The $E_{\rm em}$ evolution of two density models: $\rho_{\rm bkg}=10^{-1}$ (red), and $\rho_{\rm bkg}=10^{-3}$ (green). Similar to the example model, the non-collision cases are also plotted (dashed). There is essentially no difference between the two models.}
\label{fig:rho_bkg}
\end{figure}

The parameters for $\rho_{\rm bkg}$-varying models are listed in Table \ref{tab:rho_bkg}. Here most of the parameters are similar to the example model. However, in order to make the code stable when applying a smaller background density, we have to increase the uniform gas pressure value to $10^{-1}$. We choose two density values.
The upper panel of Figure \ref{fig:rho_bkg} shows the $E_{\rm em}$ evolution of these two models. There is no significant difference between these two models. This is understandable. Since the force balance is mainly controlled by the background gas pressure rather than density, varying the background density does not lead to significant change in the result.

\begin{table}[!htb]
\centering
\caption{The $\rho_{\rm bkg}$-varying models}
\begin{tabular}{ccccccccc}
\toprule
Model name & $\sigma_{\rm b,i}$ & $B_{\rm b,0}$ & $\alpha$ & $\left|V_{\rm b,z}\right|$ & $P$ & $\rho_{\rm bkg}$ & $z_d$ & $x_s$ \\
\midrule
Model:$\rho_{\rm bkg}$-1 & 8 & 0.8$\sqrt{4\pi}$ & 3 & 0.4c & $10^{-1}$ & $10^{-1}$ & 4.4 & 1.0\\
Model:$\rho_{\rm bkg}$-3 & 8 & 0.8$\sqrt{4\pi}$ & 3 & 0.4c & $10^{-1}$ & $10^{-3}$ & 4.4 & 1.0\\
\bottomrule
\end{tabular}
\label{tab:rho_bkg}
\end{table}

\subsection{$\alpha$ value}\label{subsec:alpha}

\begin{figure}[!htb]
\plotone{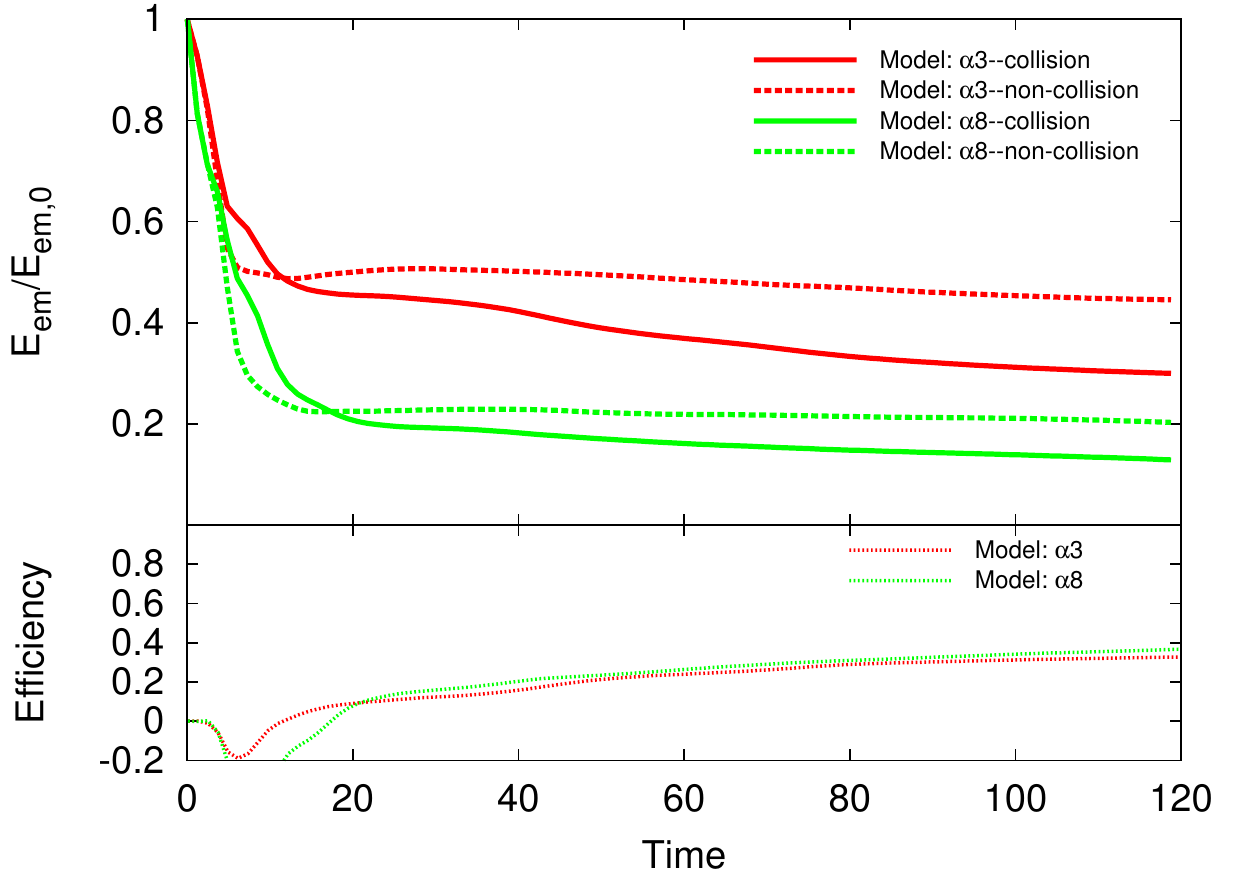}
\caption{The $E_{\rm em}$ evolution of two $\alpha$ models: $\alpha=3$ (red) and $\alpha=8$ (green). Similar to the example model, the non-collision cases for each model (dashed) are also plotted. The $E_{\rm em}$ dissipation efficiency is slightly larger for the model with a larger $\alpha$.}
\label{fig:alpha}
\end{figure}

The $\alpha$ parameter (introduced in Eq.(\ref{equ:B_phi})) defines the ratio between the toroidal and poloidal components of the initial magnetic field configuration. In the example model, we adopted $\alpha=3$, which means that the toroidal flux roughly equals to the poloidal flux. On the other hand, the central engine of GRBs (and probably AGNs as well) is likely rapidly rotating, so that the toroidal flux may be (much) larger than the poloidal flux and $\alpha > 3$. In this subsection we study the effect of $\alpha$ on the $E_{\rm em}$ energy evolution and dissipation efficiency. The parameters we used are listed in Table \ref{tab:alpha}. Here the parameters other than $\alpha$ are the same as the example model. We choose two $\alpha$ values (3 and 8). The upper panel of Figure \ref{fig:alpha} shows the $E_{\rm em}$ evolution of these two models. The initial ``self adjustment" phase due to the expansion is more significant when $\alpha$ becomes higher, since the net magnetic pressure becomes larger when $\alpha$ increases \citep{Li06}. This requires more expansion of the blobs before reaching the pressure balance with the ambient gas. The lower panel of Figure \ref{fig:alpha} shows the $E_{\rm em}$ dissipation efficiency of these two models. The model with a higher $\alpha$ value has a slightly higher efficiency. This is probably because the reconnections mainly come from the toroidal component. If $\alpha$ is much greater than 3 in realistic astrophysical systems (e.g. in GRBs and AGNs), the efficiency derived here can be regarded as a safe lower limit. 

\begin{table}[!htb]
\centering
\caption{The $\alpha$-varying models}
\begin{tabular}{ccccccccc}
\toprule
Model name & $\sigma_{\rm b,i}$ & $B_{\rm b,0}$ & $\alpha$ & $\left|V_{\rm b,z}\right|$ & $P$ & $\rho_{\rm bkg}$ & $z_d$ & $x_s$ \\
\midrule
Model:$\alpha 3$ & 8 & $\sqrt{4\pi}$ & 3 & 0.3c & $10^{-2}$ & $10^{-1}$ & 4.4 & 1.0\\
Model:$\alpha 8$ & 8 & $\sqrt{4\pi}$ & 8 & 0.3c & $10^{-2}$ & $10^{-1}$ & 4.4 & 1.0\\
\bottomrule
\end{tabular}
\label{tab:alpha}
\end{table}

\subsection{Adiabatic index}\label{subsec:adi}

\begin{figure}[!htb]
\plotone{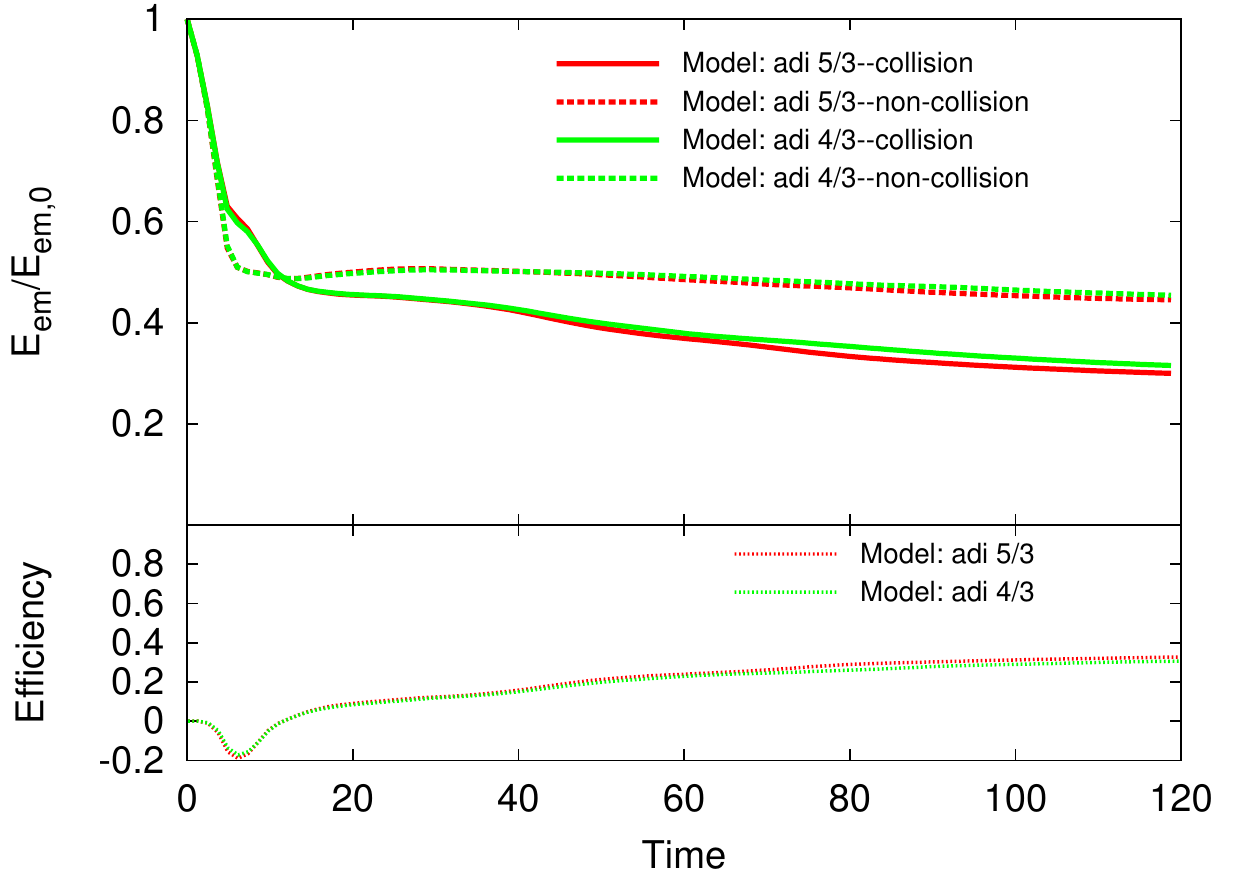}
\caption{The $E_{\rm em}$ evolution of two $\hat{\gamma}$ models: $\hat{\gamma}=5/3$ (red) and $\hat{\gamma}=4/3$ (green). The non-collision cases (dashed) for both models are also shown.
The $E_{\rm em}$ evolution and dissipation efficiency are similar for these two models.}
\label{fig:adi}
\end{figure}

Finally, in all above simulations we have used a simple uniform adiabatic index $\hat{\gamma}=5/3$, since most of the regimes are at most mildly relativistic. Nonetheless, in some high temperature regimes in the reconnection outflows, the adiabatic index may be close to the relativistic limit $\hat{\gamma}=4/3$. In principle, for a more accurate study, we need to calculate the adiabatic index between $5/3$ to $4/3$ based on the temperature distribution and time evolution cell by cell. In this subsection, based on the simplified uniform adiabatic index model, we compare the difference between these two limits: $\hat{\gamma}=5/3$ and $4/3$. We name them as ``Model:adi $5/3$" and ``Model:adi $4/3$", respectively. The parameters that we used are listed in Table \ref{tab:adi}. The ``Model:adi $5/3$" is just the example case with resolution $256^3$. The upper panel of Figure \ref{fig:adi} shows the $E_{\rm em}$ evolution of these two models. Similar to the example model, we also show the non-collision case for ``Model:adi $4/3$''. We find that there is only a slight difference between these two models. We therefore conclude that the simple uniform adiabatic index model with $\hat\gamma=5/3$ catches the essence of the collision and magnetic dissipation physics discussed in this paper.

\begin{table*}[!htb]
\centering
\caption{Two $\hat\gamma$ models}
\begin{tabular}{cccccccccc}
\toprule
Model name & $\hat{\gamma}$ & $\sigma_{\rm b,i}$ & $B_{\rm b,0}$ & $\alpha$ & $\left|V_{\rm b,z}\right|$ & $P$ & $\rho_{\rm bkg}$ & $z_d$ & $x_s$ \\
\midrule
Model:adi $5/3$ & 5/3 & 8 & $\sqrt{4\pi}$ & 3 & 0.3c & $10^{-2}$ & $10^{-1}$ & 4.4 & 1.0\\
Model:adi $4/3$ & 4/3 & 8 & $\sqrt{4\pi}$ & 3 & 0.3c & $10^{-2}$ & $10^{-1}$ & 4.4 & 1.0\\
\bottomrule
\end{tabular}
\label{tab:adi}
\end{table*}

\subsection{Summary for this section}\label{subsec:summary_para}

In this section, we have done a series of extended parameter studies. We find that the $E_{\rm em}$ dissipation efficiency is relatively insensitive to the variations of most parameters compared with the $E_{\rm em}$ evolution itself. We conclude that the two conclusions drawn in Section \S\ref{sec:example}, namely, a high collision-triggered magetic dissipation efficiency and the existence of reconnection-facilitated outflow minijets, are robust.

\section{Multiple collisions}\label{sec:4blobs}

So far we have only studied collisions between two high-$\sigma$ blobs. In reality, multiple collisions between several or even tens of blobs may occur in GRB/AGN jets, so that a much more complex configuration in the collision regions with multiple reconnection layers and outflows can be achieved (as envisaged in the ICMART model, \citealt{ZhangYan11}). Simulating multiple collisions with random blobs are technically heavy duty and require extended simulation efforts. Nonetheless, in this section, we present a preliminary four-blob interaction simulation as the first step towards a more realistic/complicated study of multiple-blob collisions. 

\begin{figure*}[!htb]
\plotone{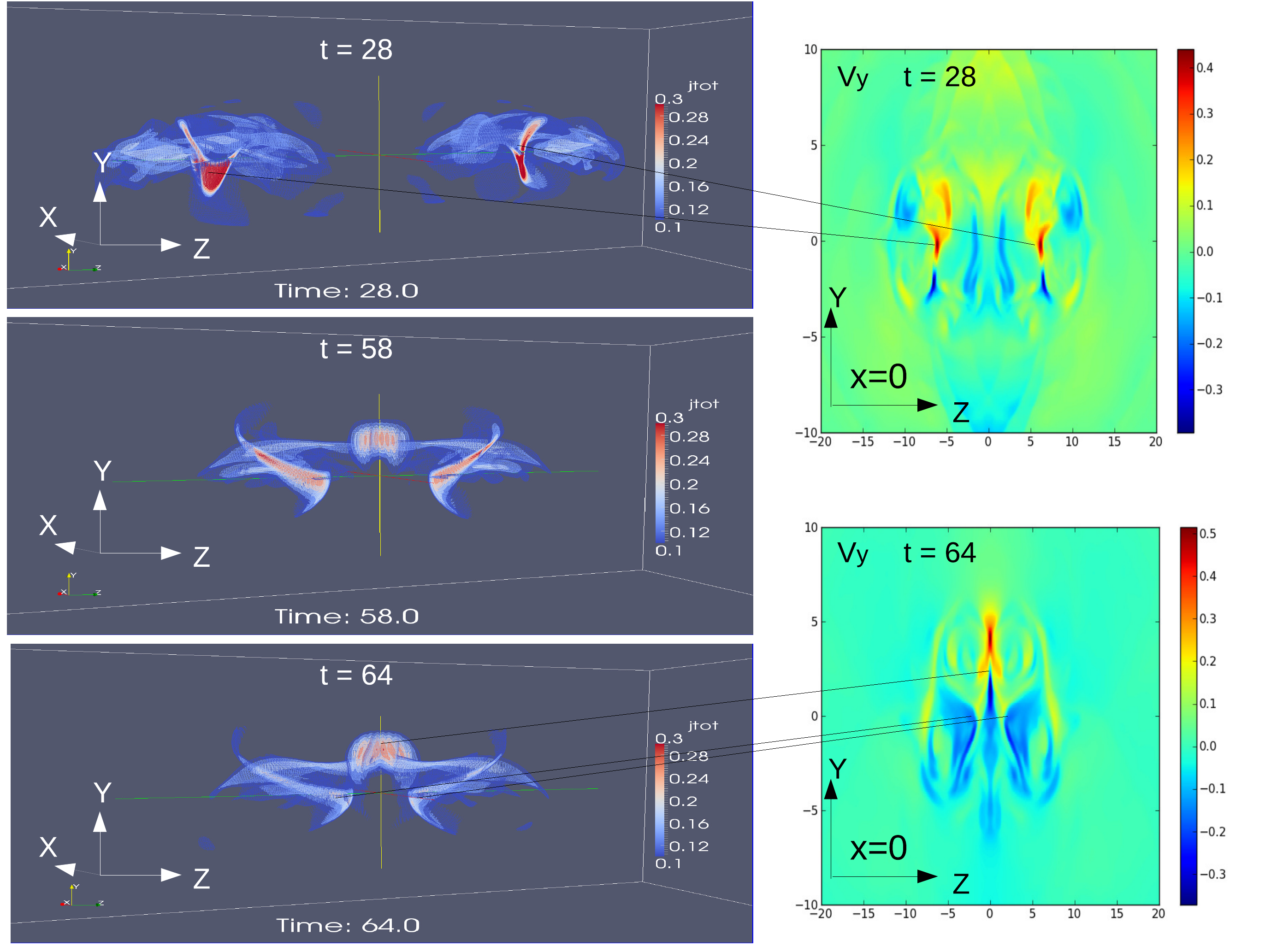}
\caption{Some results of the four-blob collision simulation. The three left panels are the 3D contour plots of the currents at $t=28$, $t=58$ and $t=64$, respectively. The two right panels are the corresponding 2D contour cuts of $V_y$ in the $YZ$ plane (x=0) at $t=28$ and $t=64$. The black lines indicate the corresponding positions of strong current layers and outflows. At least three strong reconnection layers with different directions are formed during the collision process.}
\label{fig:4blobs_current}
\end{figure*}

The size of the simulation box is $20\times 20\times 40$ with resolution $256\times 256\times 512$, which means that the grid size is the same as the example case of two blobs with resolution $256^3$ in Section \S \ref{sec:example}. The parameters used in the simulation are listed in Table \ref{tab:4blobs}. Here most of the parameters are the same as the example model. The different ones include the initial velocities of the four blobs and the distances between blobs. For the four blobs along the $+Z$ direction (bulk motion direction of the global jet) in the simulation frame (center-of-mass frame of the blobs), the velocities are $0.8c$, $0.3c$, $-0.3c$, and $-0.8c$, respectively. The initial distances between them are 5, 15, and 5, respectively, which means that the inner two blobs would collide with their nearby outer neighbors first to form two merged knots, before the two knots collide again. Figure \ref{fig:4blobs_current} shows the 3D contour plots of the current, and the corresponding 2D contour cuts of the $y$-component of the outflow velocity($V_y$) in the $YZ$ plane. At $t=28$ the first collisions between the two pairs of outside blobs form two strong reconnection regions which are similar to the example model in \S \ref{sec:example}. At later times, these two post-collision knots collide again and form a third strong reconnection region in the middle, and the original two reconnection regions also continuously evolve with time. These three reconnection regions form a more complex configuration than the collision with only two blobs as studied in \S \ref{sec:example}.

\begin{figure}[!htb]
\plotone{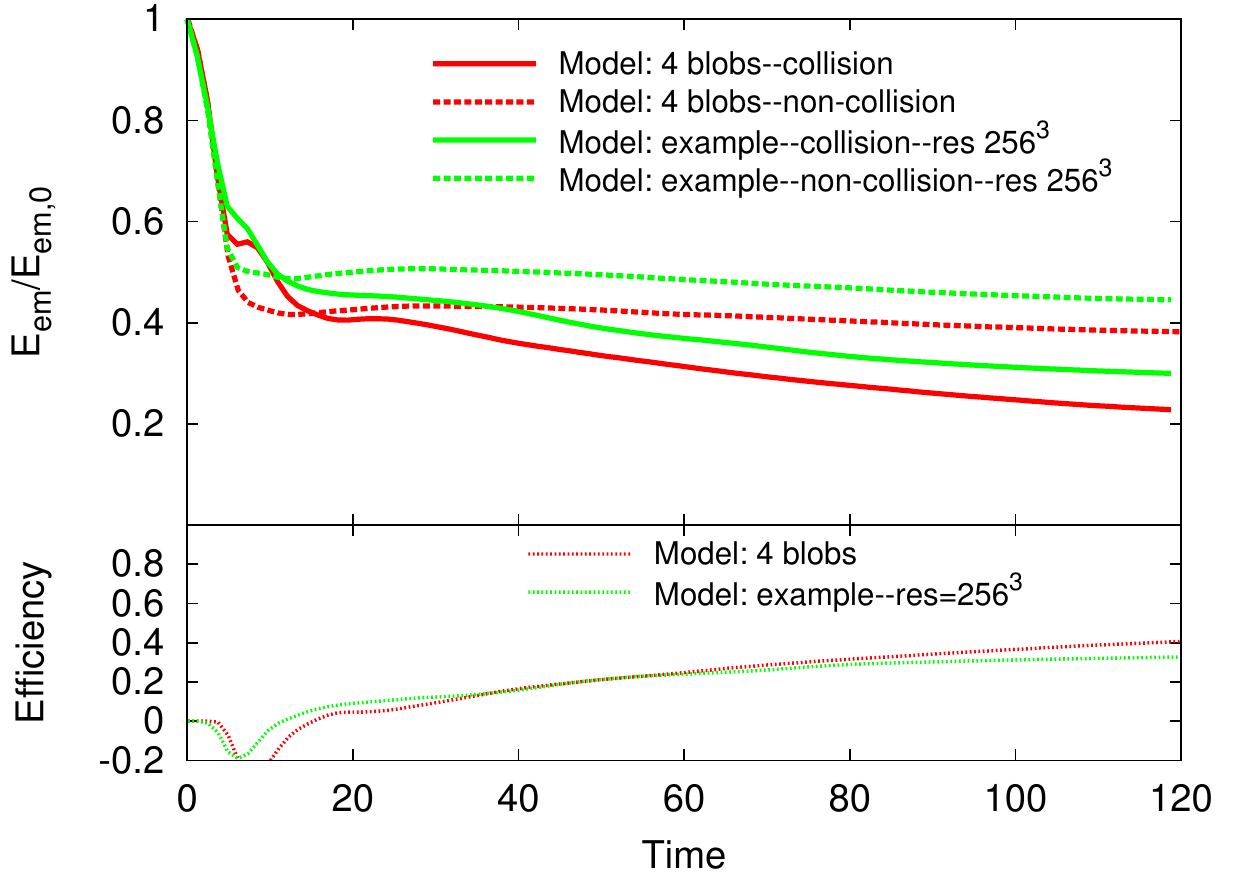}
\caption{The $E_{\rm em}$ evolution of four-blob model compared with the two-blob model at resolution $256^3$. For comparison, the non-collision model is also shown. 
The $E_{\rm em}$ dissipation efficiency in the final quasi-steady phase for the four-blob model is significantly higher than that of the two-blob model.}
\label{fig:4blobs_energy}
\end{figure}

The upper panel of Figure \ref{fig:4blobs_energy} shows the $E_{\rm em}$ evolution of this case. For comparison, we also show the non-collision case and the two-blob collision case (with same resolution), and calculate the $E_{\rm em}$ dissipation efficiency. We find that the $E_{\rm em}$ dissipation efficiency in the final quasi-steady phase is around $40\%$  (lower panel of Figure \ref{fig:4blobs_energy}), which is higher than the example two-blob case with resolution $256^3$ (lower panel of Figure \ref{fig:f_res}). This suggests that multiple collisions can facilitate further reconnection-triggered magnetic dissipations, making the system reaching a higher $E_{\rm em}$ dissipation efficiency.

\begin{table*}[!htb]
\centering
\caption{The four-blob collision model}
\begin{tabular}{ccccccccc}
\toprule
Model name & $\sigma_{\rm b,i}$ & $B_{\rm b,0}$ & $\alpha$ & $V_{\rm b,z}$ & $P$ & $\rho_{\rm bkg}$ & $z_d$ & $x_s$ \\
\midrule
Model: 4 blobs & 8 & $\sqrt{4\pi}$ & 3 & 0.8c/0.3c/-0.3c/-0.8c & $10^{-2}$ & $10^{-1}$ & 5/15/5 & 1.0\\
\bottomrule
\end{tabular}
\label{tab:4blobs}
\end{table*}

\section{Conclusions and astrophysical applications}\label{sec:summary}

In this paper, using a 3D SRMHD code, we carried out a series of simulations to study collisions between high-$\sigma$ magnetic blobs. Through a detailed example simulation and an extended parameter space survey, we have reached the following robust results:

\begin{itemize}
\item Collisions trigger significant EMF energy dissipation. Detailed analyses of the numerical data during different stages of the collision process suggest that such dissipation is facilitated by collision-induced magnetic reconnection. The efficiency of $E_{\rm em}$ dissipation in our simulations is around 35\%, which is insensitive to the numerical resolution and several initial condition parameters, such as $\sigma_{\rm b,i}$, $\Gamma_{\rm rel}$, $z_d$, $P$, and $\rho_{\rm bkg}$. It depends on the impact parameter $x_s$, which defines the area of the anti-parallel regions in the contact surface of the two blobs. As long as a small offset exists, significant dissipation is facilitated. 
\item Our simulations suggest that the collision process is essentially inelastic. Even though there is some kind of bouncing back in the early stage of the collision evelution, the strong reconnection effect in the contact surface efficiently dissipates the magnetic energy and reduces the magnetic pressure. As a result, the two high-$\sigma$ blobs merge into one larger blob with a ``$\infty$"-shaped magnetic configuration (see more details in Section \S \ref{subsec:physical}). Assuming complete inelastic collision, an estimated dissipation efficiency \citep{ZhangYan11} is found consistent with the efficiency derived from the numerical data.
\item From our simulations, we find that magnetic reconnection events can induce relativistic, multi-orientation outflows. Even in two-blob simulations, as long as an offset exists ($x_s \neq 0$), 3D outflows are formed. For four-blob collisions, significant outflows exist in three distinct regions. These outflows would mimic ``mini-jets'' as invoked in the astrophysical models of GRBs and AGNs. 
The maximum outflow velocity ($V_{\rm out}$) in our simulations is only $0.75c$. However, due to the significant resolution-dependence behavior as described in Section \S \ref{subsec:res}, we still have not reached the convergence for the outflow velocity, so that that value is only the lower limit of $V_{\rm out}$. In the simulations, we found that $\Gamma_{\rm out}$ can reach and even exceed local $\Gamma_{\rm A}$ and $\Gamma_{\rm ms}$ (\S \ref{subsec:v_out}), both are relativistic numbers if $\sigma > 1$. Also a larger $\Gamma_{\rm rel}$ tends to give a larger $V_{\rm out}$ value. In principle, with a high-resolution simulation and for a large $\sigma_{\rm b,i}$ and $\Gamma_{\rm rel}$, an even larger mini-jet Lorentz factor is achievable.
\item We found an interesting linear relationship between the initial ($\sigma_{\rm b,i}$) and the finial ($\sigma_{\rm b,f}$) values of the $\sigma$ parameter of the blob (Eq.(\ref{equ:sigmas})). 
The range of $\sigma_{\rm b,i}$ we have explored is not very large due to the code capability constraint. It is valuable to study this intriguing behavior in a larger range of $\sigma_{\rm b,i}$ in the future. 
\item Our preliminary simulations of multiple collisions among multiple high-$\sigma$ blobs suggest that the collisions would give rise to more complex configurations of the reconnection layers and multi-orientation outflows, with a higher EMF energy dissipation efficiency. This suggests that the multiple collisions of many high-$\sigma$ blobs can potentially generate many mini-jets with relatively random directions, as required by some theoretical models of astrophysical jets.
\end{itemize}

These numerical simulations have profound implications to understand astrophysical jets, such as GRBs, AGNs, X-ray binaries, Crab nebula, and so on. In the following, we discuss their direct applications to GRB and AGN models.

\subsection{GRBs}

As we mentioned in the introduction section, \cite{ZhangYan11} proposed the ICMART model to interpret the prompt emission of GRBs. This model invokes collision-induced magnetic dissipation of moderately high-$\sigma$ blobs, which is the motivation of our simulations. The ICMART model was suggested to have several salient features that can potentially interpret various observations not easy to interpret within the MDF internal shock models. Our simulations verified several assumptions/speculations adopted in the original model of \cite{ZhangYan11}.

First, \cite{ZhangYan11} claimed that ICMART processes should have a significantly higher energy dissipation efficiency than internal shocks, which is more consistent with the GRB observations \citep{panaitescu02,zhang07}. They assumed that once ICMART is triggered, the two colliding shells would merge completely in an inelastic collision. The $\sigma$ values of the two shells/blobs drop significantly from an initial value to a much lower final value. Energy and momemtum conservations suggest that the energy dissipation efficiency is high, up to 10s of percent, depending on the final $\sigma$ value of the merged blob. If $\sigma_{\rm b,f} \sim 1$, they found that the efficiency is close to 50\%. Our detailed simulations verified all these assumptions/speculations. Indeed significant magnetic dissipation occurs due to collision-induced magnetic reconnection. The collision process is essentially inelastic, and the energy dissipation efficiency is indeed high, which is $\sim 35\%$ in for two-blob collisions and $\sim 40\%$ for four-blob collisions. One surprising result is that the final value $\sigma_{\rm b,f}$ is linearly correlated with the intial value $\sigma_{\rm b,i}$ (Eq.(\ref{equ:sigmas})), so that the efficiency does not sensitively depend on $\sigma_{\rm b,f}$. More studies are needed to reveal the underlying physics of this correlation.

Second, the ICMART model invokes the central engine activities to interpret the broad pulses in the GRB light curves, but requires the existence of mini-jets to account for the rapid variability component. \cite{ZhangZhang14} used this concept to perform a series of Monte Carlo simulations and reproduced a range of highly variable light curves with both slow and fast components as seen in observational data \citep{Gao12}. The required Lorentz factor of the mini-jets is in the range of 2-15 \citep{ZhangZhang14}. In our simulations, the outflows of reconnection layers can reach mildly relativistic speed. From the orientation point of view, one major reconnection current layer between two colliding blobs already generates multi-orientation outflows (see Figure \ref{fig:stages}), in addition to systematical global rotation and twist due to the slightly initial misalignment. Furthermore, by invoking four-blob collisions, we clearly find three major reconnection layers with different directions. Each of them has their own 3D outflow systems similar to two-blob collision cases, which gives a more complex space-time distribution of the outflow directions. In realistic systems, collisions of tens of blobs would lead to more complicated 3D mini-jet structure, which would account for observed GRB light curves. It is possible in much smaller scales not resolved by the current simulations, perturbations may induce turbulent reconnections, which may make even smaller mini-jets in the current outflows. Dedicated local simulations are needed to verify or refute such a speculation. With the current global simulations, one is confident that even without turbulence, collision-induced reconnection layers can already generate large-scale mini-jets in the bulk jet of a GRB, which would give interesting variability features in the light curves.

Finally, our simulations show significant evolution of the magnetic field configuration during one ICMART event. From Figure \ref{fig:field}, we can see that during the early collision-driven reconnection phase (e.g. the ``plateau" phase), the strengths of $B_x$ and $B_y$ components decrease and the $B_z$ component increases, which significantly changes the magnetic field configuration while still keeping a relative ordered magnetic configuration in a relatively short time duration. The behavior may potentially interpret the significant change of the polarization angle during the prompt emission phase observed in GRB 100826A \citep{yonetoku11}.

\subsection{AGNs}

Some blazars show very fast TeV flares whose durations are only several minutes \citep{aharonian07,albert07}. This duration is much shorter than the light crossing time for the entire system, which means that emission comes from a small local region. The requirement of emitting TeV photons also demand a much larger Lorentz factor in the emission region (greater than 50, \citealt{begelman08,mastichiadis08}) than what is inferred for the bulk motion (typically smaller than 10, \citealt{giroletti04,piner04}). A successful model to interpret the observations is the ``jets in a jet'' model proposed by \cite{Giannios09}. This model invokes current-instability triggered local magnetic reconnections in a global, Poynting-flux-dominated jet. These local reconnections generate local outflows or mini-jets with a comoving Lorentz factor around a few. Our simulations give an alternative process to trigger the local reconnections by considering ICMART events, i.e. collisions among magnetic knots/blobs inside the global jet. Since the knots in AGNs have already been observed \citep{marscher02,chat09,doi2011}, the collisions would very likely happen, which trigger the local reconnections and generate the mini-jets as needed in their model.

We can also roughly estimate the time scale using our simulation results and the parameters in the model of \cite{Giannios09}. They estimated that the typical size of the blob in the rest frame of the blob is around $10^{14} {\rm cm}$. Since the Lorentz factor of the blob in the comoving frame of the global jet is equal to 10 (assuming $\sigma=100$), the size of the global jet in the comoving frame is about $10^{13} {\rm cm}$, which can be treated as $L_0$ in our Table \ref{tab:t2}. Thus one time unit in our simulations can be normalized as $t_0=L_0/c \sim 3\times 10^2 {\rm s}$. The duration of the reconnection-facilitated energy dissipation is about 50 time units in our example case (see Figure \ref{fig:f1} from t=30 to t=80), which can be translated to about $10^4 {\rm s}$ in the rest frame of the global jet. The duration in the observer frame is $\sim 10^4 {\rm s}/\Gamma_j \sim 10^4 {\rm s}/10 = 10^3 {\rm s}$, which is very close to the observed durations of the flares.

\begin{acknowledgements}
This work is supported by the LANL/LDRD program and Institutional Computing Programs at LANL and
by DOE/Office of Fusion Energy Science through CMSO, and by NASA through grant NNX14AF85G funded to UNLV.  We thank helpful discussion and suggestions from Fan Guo, Xiaoyue Guan, Jim Stone, Feng Yuan, and Donald Lamb.
\end{acknowledgements}

\end{CJK*}
\end{document}